\documentclass[11pt,twoside]{atmp}

\usepackage{amsmath,amssymb}
\usepackage{fullpage,times,epsf}
\usepackage{graphics}
\usepackage{graphicx}
\usepackage{caption}
\usepackage{float}
\usepackage[square,numbers,sort&compress]{natbib}
\usepackage[dvipdfm]{hyperref}
\usepackage[all]{xy}
\usepackage{color}

\allowdisplaybreaks

\def\beginproof{\par\noindent\textbf{Proof.}~~}
\def\endproof{\ \vbox{\hrule\hbox{\vrule height1.0ex\hskip1.0ex\vrule}\hrule }\par\medskip}

\newtheorem{theorem}{Theorem}[section]

\newtheorem{lemma}[theorem]{Lemma}

\newtheorem{proposition}[theorem]{Proposition}

\newtheorem{remark}[theorem]{Remark}
\newcommand{\me}{\mathrm{e}}
\newcommand{\mi}{\mathrm{i}}
\def\diag{\mathrm{diag}}
\def\lam{\lambda}
\def\mb{\mathbf}
\def\ra{\rightarrow}
\def\tnn{\mathrm{tnn}}
\def\Gr{\mathop{\rm Gr}\nolimits}
\def\GrNMtnn{\Gr^\tnn_{N\!M}}

\def\lim{\mathrm{lim}}
\def\sech{\mathrm{sech}}
\def\rank{\mathrm{rank}}

\def\deg{\mathrm{deg}}
\def\coeff{\mathrm{coeff}}
\def\det{\mathrm{det}}

\def\eref#1{(\ref{#1})}
\begin{document}

\title
{Multi-component Wronskian solution to the Kadomtsev-Petviashvili equation}


\author
{Tao Xu$^{1}$,
Fu-Wei Sun$^{2}$, Yi Zhang$^{1}$  and Juan Li$^{3,4}$}
\address{
1. College of Science,  China University of Petroleum, Beijing 102249, China. \\[1mm]
2. College of Science, North China University of Technology, Beijing 100041, China. \\[1mm]
3. State Key Laboratory of Remote Sensing Science, Jointly Sponsored \\[1mm]
by the Institute of Remote Sensing Applications of Chinese of Sciences \\[1mm]
Academy and Beijing Normal University, Beijing 100101, China.\\[1mm]
4. Demonstration Centre, Spaceborne  Remote Sensing National Space \\[1mm]
Administration, Beijing 100101, China.
}  
\addressemail{E-mail address: xutodd@126.com (T. Xu)}

\begin{abstract}
It is known that the Kadomtsev-Petviashvili (KP) equation can be decomposed into the first two members of the coupled Ablowitz-Kaup-Newell-Segur (AKNS) hierarchy by the binary nonlinearization of Lax pairs. In this paper, we construct the $N$-th iterated Darboux transformation (DT) for the second- and third-order $m$-coupled AKNS systems. By using together the $N$-th iterated DT  and  Cramer's rule, we find  that the KPII equation has the unreduced multi-component Wronskian solution and the KPI equation admits a reduced multi-component Wronskian solution. In particular, based on the unreduced and reduced two-component Wronskians, we obtain two families of fully-resonant line-soliton solutions which contain arbitrary numbers of asymptotic solitons as $y \ra \mp \infty$ to the KPII equation, and the ordinary $N$-soliton solution to the KPI equation. In addition, we  find that  the KPI line solitons propagating in parallel can exhibit the bound state at the moment of collision.
\end{abstract}

\maketitle

\cutpage 

\setcounter{page}{2}

\noindent

\section{Introduction} \label{introduction}

In 1970, Kadomtsev and Petviashvili~\cite{Kadomtsev1970} derived the following (2+1)-dimensional nonlinear dispersive wave equation
\begin{equation}
 (-4u_t + 6 u\,u_x + u_{xxx})_x +3\,\sigma^2 u_{yy} =0  \quad (\sigma^2=\pm 1), \label{KP}
\end{equation}
to study the stability of soliton solutions of the Korteweg-de Vries (KdV) equation with respect to weak transverse perturbations. Eq.~\eref{KP} is usually called the  Kadomtsev-Petviashvili (KP) equation, where $\sigma^2=- 1$ and $1$ represent the KPI and KPII equations, respectively.
As a natural extension of the KdV equation in two dimensions,
the KP equation~\eref{KP} with both cases of $\sigma^2$ has arisen in various physical contexts, e.g., plasma physics~\cite{Kadomtsev1970}, surface and internal
water waves~\cite{Ablowitz1979}, 
etc.
The KP equation~\eref{KP} is a universal completely-integrable (2+1)-dimensional  nonlinear evolution equation (NLEE)~\cite{Ablowitz1992} and its many integrable properties have been revealed in the past forty years, including the Lax representation~\cite{Dryuma1974}, lump solutions to the KPI equation~\cite{Satsuma1979}, line-soliton solutions to the KPI and KPII equations~\cite{Satsuma1976},
an infinite number of symmetries and conservation laws~\cite{Oevel1982},
Painlev\'{e} property~\cite{Weiss1983}, B\"{a}cklund transformation~\cite{Chen1983b},
Darboux transformation (DT)~\cite{Matveev1979}, a recursion operator and bi-Hamiltonian structure~\cite{Fokas1986,Fokas1988},  and similarity reductions to the Painlev\'{e}-type ordinary differential equations~\cite{David1986} and integrable (1+1)-dimensional NLEEs~\cite{Lou1990}.

In recent years, the KPII equation has attracted intensive attention~\cite{Biondini2003,Biondini2006,Biondini2007PRL,Chak2008,Chak2009,
Kodama2009JPA,Kodama2010JPA,Kodama2011}  because
a large variety of soliton solutions of this equation were overlook in the literature before 2003. It is known that the  KPII equation admits the solution $u = 2\frac{\partial^2}{\partial x^2}  \ln \tau(x,y,t) $ with the tau function $\tau (x,y,t)$ expressible in the Wronskian form~\cite{Freeman1983PLA}
\begin{align}
\tau(x,y,t)  =
\begin{vmatrix}
f_1&f_2&\cdots&f_N\\
\frac{\partial f_1}{\partial x} & \frac{\partial f_2}{\partial x} &\cdots & \frac{\partial f_N}{\partial x}
\\\vdots&\vdots&\ddots&\vdots\\
\frac{\partial^{N-1}\!  f_1}{\partial x^{N-1} } & \frac{\partial^{N-1} \! f_2}{\partial x^{N-1} } &\cdots & \frac{\partial^{N-1}\! f _N}{\partial x^{N-1}}\end{vmatrix},
\label{KPSoliton}
\end{align}
where 
$ \{ f_h \}_{h=1}^N $ are a set of linearly-independent solutions of the
linear system: $\sigma f_y=  f_{xx}$ ($\sigma = \pm 1$), $f_t= f_{xxx}$.
Ref.~\cite{Biondini2006} has  revealed a general family of multi-soliton solutions to the KPII equation  by noting that the functions $ \{
f_h \}_{h=1}^N $ can be chosen as
$f_h(x,y,t) = \sum_{l=1}^M a_{hl}\me^{\kappa_l x + \sigma\kappa_l^2 y + \kappa_l^3 t}$ ($1\leq h \leq N$),
where $M$ is an positive integer greater than $N $, the phase parameters $\{\kappa_l\}_{l=1}^M$ are distinct, the constant coefficients $a_{hl}$ $(1\leq h\leq N; 1\leq l\leq M)$ define an $N\times M$ coefficient matrix
$A:=(a_{hl})$, which is required to be of full rank  [i.e., $\rank(A)=N]$
and all of whose nonzero $N\times N$ minors must be sign definite.

If the matrix $A$ is in the reduced row-echelon form (RREF)  and satisfies the
\emph{irreducibility conditions} that each column of $A$ contains at least one nonzero element and each row of
$A$ contains at least one nonzero element in addition to the pivot (first nonzero) entry,
the function $\tau$ in Eq.~\eref{KPSoliton} can yield a multi-soliton configuration
with $N_+= N $ asymptotic  solitons as $y\ra +\infty$ and
$N_{-}=M-N$ asymptotic solitons as $y\ra -\infty$~\cite{Biondini2006}.
Moreover, it has been indicated~\cite{Biondini2007PRL,Chak2008,BiondiniXu2012,Chak2009,Kodama2010JPA} that the classification problem of soliton solutions to the KPII equation can be solved from the viewpoint of algebraic geometry  and combinatorics because each equivalence class of $(N, M-N)$-soliton solutions corresponds to a derangement (which is a permutation with no fixed point) on $\{1,2\dots, M\}$ with $N$ excedances~\cite{Chak2008,Chak2009}, and
each RREF $N\times M$ matrix $A$ (all of whose maximal minors are non-negative)
belongs to a unique totally non-negative (TNN) Grassmann cell which is
a further decomposed cell of the Schubert cell of the
TNN  Grassmannian $\GrNMtnn$~\cite{BiondiniXu2012,Postnikov2006}.
In addition, the initial value problem of the KPII equation with V- and X-shape
initial waves consisting of two distinct line solitons has also been studied by the direct numeric simulation~\cite{Chak2009,Kodama2009JPA,Kodama2010JPA,Kodama2011},
which gives an explanation of the Mach reflection phenomena in shallow water~\cite{Chak2009,Kodama2010JPA}.


Unlike the work in Refs.~\cite{Biondini2003,Biondini2006,Chak2008,Chak2009,
Kodama2009JPA,Kodama2010JPA,Kodama2011,Biondini2007PRL}, the present paper is going to construct
the multi-component Wronskian solution to the KP equation~\eref{KP} and
 explore the underlying soliton solutions.
The motivation of this study comes from the
observation in two aspects. First,  the binary nonlinearization of two symmetric Lax pairs can decompose the KP equation~\eref{KP} into the first two members of the coupled Ablowitz-Kaup-Newell-Segur (AKNS) hierarchy~\cite{NonlKP1,NonlKP2}:
\begin{subequations}
\begin{align}
&  p_{j,y} = - \sigma^{-1} \bigg(
p_{j,xx} - 2 \sum_{n = 1}^m\,p_n\,q_n\, p_j \bigg) \quad (1\leq j\leq m), \\
&
q_{j,y} = \sigma^{-1}\bigg( q_{j,xx} - 2 \sum_{n =
1}^m\,q_n\,p_n\, q_{j}\bigg) \quad  (1\leq j\leq m),
\end{align} \label{GNLS}
\end{subequations}
and
\begin{subequations}
\begin{align}
&  p_{j,t} = p_{j,xxx} - 3 \sum_{n = 1}^mp_{n}\,q_{n}\,
p_{j,x} - 3 \sum_{n = 1}^m p_{n,x}\,q_{n}p_j   \quad
(1\leq j\leq m),
\\
& q_{j,t} =  q_{j,xxx} - 3 \sum_{n = 1}^mq_{n}\,p_{n}\,
q_{j,x} - 3 \sum_{n = 1}^mq_{n,x}\,p_{n} q_j= 0  \quad
(1\leq j\leq m),
\end{align}  \label{GMKdV}
\end{subequations}
where $m$ is an arbitrary positive integer, Systems~\eref{GNLS} and~\eref{GMKdV} are respectively referred to as the second- and third-order $m$-coupled AKNS systems, and
\begin{align}
u= -2\sum_{j=1}^m p_j q_j,  \label{KPAKNS}
\end{align}
exactly satisfies the KP equation~\eref{KP}. Second, the soliton solutions to both the vector nonlinear Schr\"{o}dinger (NLS) equation and vector complex modified KdV equation, which are respectively  two reductions of Systems~\eref{GNLS} and~\eref{GMKdV}, have been found to be expressible in terms of the multi-component Wronskian~\cite{NWrVNLS1,NWrVNLS2,NWrVH}. Therefore, it is reasonable to infer that the KP equation~\eref{KP}  also admits the multi-component Wronskian solution.  The structure of this paper is organized as follows.

In Section 2, we  follow the way in Ref.~\cite{GHZ} to explicitly construct the $N$-th iterated DT for Systems~\eref{GNLS} and~\eref{GMKdV}, which provides a basis for us to further obtain the multi-component Wronskian solution to the KP equation~\eref{KP}.  We note that Ref.~\cite{GHZ} has presented a general scheme of constructing the $N$-th iterated DT for the AKNS hierarchy. However, there has been an absence of rigorous proof for such $N$-th iterated DT. In this section, we  give a complete proof that the Lax pairs associated with Systems~\eref{GNLS} and~\eref{GMKdV} keep the form-invariance under the $N$-th iterated DT.


In most of the previous literature (see, for example,
Refs.~\cite{Freeman1983PLA,DWrAKNSKN}),  researchers usually obtain the Wronskian solutions by assuming that a given NLEE admits the Wronskian solution with certain condition, and then proving their conjecture by direct substitution the solution into the original equation or bilinear equation(s). In fact, one can also realize the Wronskian solutions to those  Lax-integrable NLEEs by using together the $N$-th  iterated DT algorithm and Cramer's rule rather than relying on the guesswork~\cite{NWrVH}. Moreover, if the Lax pair has been proved to be form-invariant under the DT, there is no need to verify the Wronskian solution again by means of the Pl\"{u}cker relation. In Section 3, we   derive the multi-component Wronskian solution to the KP equation~\eref{KP} by combining the $N$-th iterated DT and Cramer's rule.

In Section 4, we study the soliton solutions to the KP equation generated by the multi-component Wronskian.  In Section 4.1,   we from an unreduced two-component Wronskian derive two families of fully-resonant line-soliton solutions to the KPII equation, which are shown to be two special cases of the soliton solutions generated by the Wronskian~\eref{KPSoliton}. In Section 4.2, based on a reduced two-component Wronskian, we obtain the ordinary $N$-soliton solution to the KPI equation which, in general, describes the elastic  collisions of line solitons. In particular, we find the bound state exhibited by the line solitons propagating in parallel  at the moment of collision.

Finally, in Section 5, we  address the conclusions of this paper.

\section{$N$-th iterated Darboux transformation}
\label{NDT}

In the scheme of the $(m+1)\times (m+1)$-matrix AKNS scattering formulation~\cite{Ablowitz1975JMP},
the Lax representation of Systems~\eref{GNLS} and~\eref{GMKdV} can be
written  in the form
\begin{subequations}
\begin{align}
& \Phi_{x}=U(\lam)\Phi = (\lam\,U_0 +U_1)\,\Phi,   \label{LaxZua} \\
& \Phi_{y}=V(\lam)\Phi = (\lam^2 V_0 +
\lam\,V_1 + V_2)\,\Phi, \label{LaxZub}  \\
& \Phi_{t}=W(\lam)\Phi= (\lam^3 W_0 +
\lam^2 W_1 + \lam W_2 +W_3)\,\Phi,  \label{LaxZuc}
\end {align}  \label{LaxZu}
\end{subequations}
with
\begin{eqnarray}
& & \nonumber
U_0=
\begin{pmatrix}
1  & \mb{0}  \\
\mb{0} & -  E_m  \\
\end{pmatrix}, \quad
U_1=\begin{pmatrix}
0  & \mb{p} \\
\mb{q}^\top & \mb{0} \\
\end{pmatrix}, \\[1mm]
& & \nonumber
V_0=-\frac{2}{\sigma}\,U_0, \quad V_1=-\frac{2}{\sigma}\,U_1, \quad
V_2=\frac{1}{\sigma}
\begin{pmatrix}
\mb{p} \mb{q}^\top  & -\mb{p}_{x}    \\[1mm]
\mb{q}^\top_{x}  & -\mb{q}^\top \mb{p}
\end{pmatrix}, \\[2mm]
& & \nonumber W_0 = 4\,\,U_0, \quad  W_1= 4\,\,U_1, \quad W_2= -2\,\sigma V_2, \\[2mm]
& & \nonumber
W_3 =
\begin{pmatrix}
\mb{p}\mb{q}_x^\top - \mb{p}_x \mb{q}^\top   & \mb{p}_{xx}  - 2\,\mb{p}\mb{q}^\top\mb{p} \\[1mm]
\mb{q}^\top_{xx} - 2\,\mb{q}^\top\mb{p}\mb{q}^\top  & \mb{q}^\top\mb{p}_x- \mb{q}_x^\top\mb{p}
\end{pmatrix},
\end{eqnarray}
where $\lam$ is the spectral parameter, $\Phi=(\phi_1, \phi_2, \dots, \phi_{m+1})^\top$ ($\top$ denotes the transpose of  vector) is the vector eigenfunction,
$\mb{p}=(p_1,p_2,\dots,p_m)$ and $\mb{q}=(q_1,q_2,\dots,q_m)$ are the vector potentials, $E_m$ is the $m\times m$ identity matrix,
and the compatibility conditions $\Phi_{xy} =\Phi_{yx}$ and $\Phi_{xt} =\Phi_{tx}$ yield Systems~\eref{GNLS} and~\eref{GMKdV}, respectively.

According to the idea of the DT~\cite{GHZ}, it requires that under the $N$-th iterated eigenfunction transformation
\begin{align}
\Phi'= T (\lam)\Phi,
\label{DT}
\end{align}
the new eigenfunction $\Phi'=(\phi'_1, \phi'_2, \dots, \phi'_{m+1})^\top $ also satisfies System~\eref{LaxZu} with the matrices $U(\lam)$, $V(\lam)$ and $W(\lam)$ replaced respectively by $U'(\lam)$, $V'(\lam)$ and $W'(\lam)$ in which the new potentials $\mb{p}' = (p'_1,p'_2,\dots,p'_m)$ and $\mb{q}'=(q'_1,q'_2,\dots,$ $q'_m)$ are respectively instead of the old potentials $\mb{p}$ and $\mb{q}$, that is,
\begin{align}
\Phi'_x=U'(\lam)\Phi', \quad \Phi'_y=V'(\lam)\Phi', \quad \Phi'_t=W'(\lam)\Phi' .
\label{NewLaxZu}
\end{align}
Thus, the Darboux matrix $T(\lam)$ has to obey the following three conditions:
\begin{subequations}
\begin{align}
& T_x(\lam) + T(\lam)U(\lam) = U'(\lam)T(\lam), \label{DTinvariancea}\\
& T_y(\lam) + T(\lam)V(\lam) = V'(\lam)T(\lam), \label{DTinvarianceb} \\
& T_t(\lam) + T(\lam)W(\lam) = W'(\lam)T(\lam), \label{DTinvariancec}
\end{align}
\end{subequations}
where Conditions~\eref{DTinvariancea}--\eref{DTinvariancec} respectively correspond to the
form-invariance of Eqs.~\eref{LaxZua}--\eref{LaxZuc}.


For the satisfaction of Conditions~\eref{DTinvariancea}--\eref{DTinvariancec}, we assume that the matrix $T(\lam)$ is of the form~\cite{GHZ}
\begin{align}
T(\lam)=
\begin{pmatrix}
\alpha(\lam) &   \beta_1(\lam) & \cdots & \beta_m(\lam) \\
\gamma_1(\lam)  &  \delta_{11}(\lam) & \cdots &  \delta_{1m}(\lam) \\
\vdots & \vdots & \ddots & \vdots  \\
\gamma_m(\lam)  &  \delta_{m1}(\lam) & \cdots &  \delta_{mm}(\lam)
\end{pmatrix}, \label{DarbouMatrix}
\end{align}
with
\begin{eqnarray}
& &  \alpha(\lam)=  \lam^N-\sum_{n=0}^{N-1}\alpha^{(n)} \lam^n, \quad \beta_j(\lam) =\sum_{n=0}^{N-1}\beta_j^{(n)}(-\lam)^n \quad (1\leq j\leq m),
 \label{DTABCD1}\\
& & \gamma_i(\lam) = -\sum_{n=0}^{N-1}\gamma_i^{(n)}\lam^n \quad (1\leq i\leq m), \quad
\delta_{ii}(\lam) = \lam^N + \sum_{n=0}^{N-1}\delta_{ii}^{(n)}(-\lam)^n \quad (1\leq i\leq m),
\label{DTABCD2}
\\
& &   \delta_{ij}(\lam) = \sum_{n=0}^{N-1}\delta_{ij}^{(n)}(-\lam)^n \quad (1\leq i, j\leq m;\, i\neq j), \label{DTABCD3}
\end{eqnarray}
where  $\beta_j^{(n)}$, $\gamma_i^{(n)}$ and $\delta_{ij}^{(n)}$ ($1\leq i, j\leq m$;  $0\leq n\leq N-1$)
are the functions of $x$, $y$ and $t$ which are determined by
\begin{align}
T(\lam_k)\Phi_k=\mb{0} \quad (1\leq k\leq K=mN+N),  \label{DTLS}
\end{align}
with $\Phi_k=(f_k, g^{(1)}_k, \dots, g^{(m)}_k)^\top$ as the solution of  System~\eref{LaxZu} corresponding to $\lam=\lam_k$ ($\lam_k\neq \lam_l$; $1\leq k, l\leq K$).
It should be noted that $\{\Phi_k\}_{k=1}^K$ are a set of linearly-independent solutions of System~\eref{LaxZu} because $\lam_k\neq \lam_l$. Accordingly, one can \emph{uniquely} determine $\beta_j^{(n)}$, $\gamma_i^{(n)}$ and $\delta_{ij}^{(n)}$ ($1\leq i, j\leq m$;  $0\leq n\leq N-1$) from~\eref{DTLS} which can be expanded as
\begin{align}
\alpha(\lam_k)     + \sum_{j=1}^m \varpi^{(j)}_k \beta_{j}(\lam_k) = 0, \quad  \gamma_i(\lam_k)
+ \sum_{j=1}^m \varpi^{(j)}_k \delta_{ij}(\lam_k) = 0 \quad (1\leq i \leq m; \, 1\leq k\leq K),
\label{UndetCoeffa}
\end{align}
where $\varpi^{(j)}_k=g^{(j)}_k/f_k  $  $(1\leq j \leq m; \, 1\leq k\leq K)$.

Before proving that Conditions~\eref{DTinvariancea}--\eref{DTinvariancec} are satisfied with the Darboux matrix $T(\lam)$ given by Eq.~\eref{DarbouMatrix}, in which $\beta_j^{(n)}$, $\gamma_i^{(n)}$ and $\delta_{ij}^{(n)}$ ($1\leq i, j\leq m$;  $0\leq n\leq N-1$) are  determined by Eqs.~\eref{UndetCoeffa}, we present the following two important lemmas.

\begin{lemma} \label{L:DTPolynomial}Let the Darboux matrix $T(\lam)$ be in the form of~\eref{DarbouMatrix} with 
$\beta_j^{(n)}$, $\gamma_i^{(n)}$ and $\delta_{ij}^{(n)}$ ($1\leq i, j\leq m$;  $0\leq n\leq N-1$)  determined by Eqs.~\eref{UndetCoeffa}.
Then, the determinant of $T(\lam)$  can be expanded as
\begin{align}
\det\,T(\lam) =\prod_{k=1}^{K}(\lam-\lam_k),  \label{TPolynomial}
\end{align}
where $\lam_k$ ($1\leq k\leq K$) are the roots of $\det\, T(\lam)$.
\end{lemma}

\beginproof
It is easy to know that $\det T(\lam) $ is a monic polynomial of degree $K$. On the other hand, one can employ  Eqs.~\eref{UndetCoeffa} to express  the first column of $T(\lam_k)$  as the linear combination of other columns:
 \begin{align}
[\alpha(\lam_k),   \gamma_1(\lam_k), \dots, \gamma_m(\lam_k)]^\top=-
\sum_{j=1}^{m}\varpi^{(j)}_k[
\beta_j(\lam_k),   \delta_{1j}(\lam_k), \dots, \delta_{mj}(\lam_k)]^\top \quad (1\leq k \leq K),
\end{align}
which suggests that $ \lam_k $ ($1\leq k \leq K$) are the roots of $\det\,T(\lam)$. Therefore, the determinant of $ T(\lam)$ can be expressed in the form of~\eref{TPolynomial}.   \endproof

\begin{lemma} \label{L:uproperties}
Let us define that
\begin{subequations}
\begin{align}
& [u_{hl}(\lam)]_{(m+1)\times (m+1)} = [T_x(\lam)+T(\lam)U(\lam)]T^*(\lam), \\
& [v_{hl}(\lam)]_{(m+1)\times (m+1)} = [T_y(\lam)+T(\lam)V(\lam)]T^*(\lam), \\
& [w_{hl}(\lam)]_{(m+1)\times (m+1)} = [T_t(\lam)+T(\lam)W(\lam)]T^*(\lam),
\end{align}
\end{subequations}
where $T^*(\lam)$ is the adjoint matrix of $T(\lam)$. Then,  $\lam_k$ ($1\leq k\leq K$) are the roots of $u_{hl}(\lam)$, $v_{hl}(\lam)$ and $w_{hl}(\lam)$ ($1\leq h,l\leq m+1$), that is,
\begin{align}
u_{hl}(\lam_k)=0, \quad  v_{hl}(\lam_k)=0, \quad  w_{hl}(\lam_k)=0
\quad (1\leq h,l\leq m+1; \, 1\leq k\leq K).
\end{align}
\end{lemma}

The proof of Lemma~\ref{L:uproperties} is given in Appendix~A. In the following, we use $\deg[f(\lam)]$ to represent the degree of the polynomial $f(\lam)$, and $\alpha^*(\lam)$,  $\beta^*_j(\lam)$, $\gamma^*_i(\lam)$ and $\delta^*_{ij}(\lam)$  to denote the algebraic cofactors of $\alpha(\lam)$, $\beta_j(\lam)$,  $\gamma_i(\lam)$ and $\delta_{ij}(\lam)$ ($1\leq i, j\leq m$), respectively.

\begin{proposition}  \label{P:xinvariance}
Assume that  $\Phi_k=(f_k, g^{(1)}_{k},  \dots, g^{(m)}_{k})^\top$  satisfies  Eq.~\eref{LaxZua} with $\lam=\lam_k$, where $1\leq k\leq K$. Then, the Darboux matrix $T(\lam)$ given by~\eref{DarbouMatrix}  obeys Condition~\eref{DTinvariancea}, provided that  $\beta_j^{(n)}$, $\gamma_i^{(n)}$ and $\delta_{ij}^{(n)}$ ($1\leq i, j\leq m$;  $0\leq n\leq N-1$) are  determined by Eqs.~\eref{UndetCoeffa}, and the $N$-th iterated
potential transformations are given by
\begin{align}
\mb{p}'=\mb{p}-2\,(-1)^{N-1}\mb{b}^{(N-1)}, \quad  \mb{q}'=\mb{q}-2\,\mb{c}^{(N-1)}, \label{PotentialTran}
\end{align}
where $\mb{b}^{(N-1)} =\big(\beta_1^{(N-1)}, \dots, \beta_m^{(N-1)} \big)$ and $\mb{c}^{(N-1)} =\big(\gamma_1^{(N-1)}, \dots, \gamma_m^{(N-1)}\big)$.
\end{proposition}

\beginproof
Let $\Pi (\lam) 
= [T_x(\lam)+T(\lam)U(\lam)]T^*(\lam)/\det\,T(\lam) $.
The calculation of algebraic cofactors $\alpha^*(\lam)$, $\beta^*_j(\lam)$,  $\gamma^*_i(\lam)$ and $\delta_{ij}^*(\lam)$ yields that $\deg[\alpha^*(\lam)]=\deg[\delta_{ii}^*(\lam)]=m N$ and  $\deg[\beta_{j}^*(\lam)]= \deg[\gamma_{i}^*(\lam)]= \deg[\delta_{ij}^*(\lam)] = mN-1 $ ($1\leq i, j \leq m$; $i\neq j$), which tells us that $\deg[u_{hh}(\lam)] =K+1 $ and  $\deg[u_{hl}(\lam)]=K$ ($1\leq h,l \leq m+1$; $h\neq l$). On the other hand, Lemmas~\ref{L:DTPolynomial} and~\ref{L:uproperties} imply that  $u_{hl}(\lam)$ ($1\leq h,l\leq m+1$) can be exactly divided by $\det\,T(\lam)$. Therefore, the matrix $\Pi (\lam)$ can be written in the form
\begin{align}
\Pi(\lam)=
\begin{pmatrix}
\lam\, \pi^{(1)}_{11} + \pi^{(0)}_{11}  & \pi^{(0)}_{12} & \dots & \pi^{(0)}_{1,m+1} \\
\pi^{(0)}_{21}  & \lam\, \pi^{(1)}_{22} + \pi^{(0)}_{22} & \dots & \pi^{(0)}_{2,m+1} \\
\vdots & \vdots & \ddots & \vdots \\
\pi^{(0)}_{m+1,1}  &  \pi^{(0)}_{m+1,2} & \dots &  \lam\, \pi^{(1)}_{m+1,m+1} + \pi^{(0)}_{m+1,m+1}
\end{pmatrix},
\end{align}
where $\pi^{(1)}_{hh}$ and $\pi^{(0)}_{hl}$ ($1\leq h,l \leq m+1$) are all the functions dependent on $x$, $y$ and $t$.
By expanding $T_x(\lam)+T(\lam)U(\lam) =\Pi (\lam) T(\lam) $ and comparing the coefficients of $\lam^{N+1}$ and $\lam^N$, we can find
\begin{eqnarray}
\nonumber & & \pi^{(1)}_{11} =1, \quad  \pi^{(0)}_{11} = 0, \quad \pi^{(1)}_{i+1,i+1} =-1, \quad \pi^{(0)}_{i+1,j+1}=0 \quad  (1\leq i,j \leq m), \\
\nonumber & & \pi^{(0)}_{1,j+1} =p_j -2\,(-1)^{N-1}\beta_j^{(N-1)}, \quad
\pi^{(0)}_{i+1,1} =q_i -2\,\gamma_i^{(N-1)}  \quad  (1\leq i,j \leq m).
\end{eqnarray}
It is obvious that  the matrix $\Pi(\lam)$ is exactly equal to $U'(\lam) $ with the new potentials $\mb{p}'$ and $\mb{q}'$ given by~\eref{PotentialTran},
which means that $T_x(\lam) + T(\lam)U(\lam) = U'(\lam)T(\lam)$.
 \endproof

Similarly, we present  other two propositions, as stated in the following:

\begin{proposition}  \label{P:yinvariance}
Assume that $\Phi_k=(f_k, g^{(1)}_{k},  \dots, g^{(m)}_{k})^\top$  satisfies  Eq.~\eref{LaxZub} with $\lam=\lam_k$, where $1\leq k\leq K$. Then, the Darboux matrix $T(\lam)$ in~\eref{DarbouMatrix}  obeys Condition~\eref{DTinvarianceb}, provided that  $\beta_j^{(n)}$, $\gamma_i^{(n)}$ and $\delta_{ij}^{(n)}$ ($1\leq i, j\leq m$;  $0\leq n\leq N-1$) are  determined by Eqs.~\eref{UndetCoeffa},  and  the new potentials $\mb{p}'$ and $\mb{q}'$ are given by~\eref{PotentialTran}.
\end{proposition}

\begin{proposition}  \label{P:tinvariance}
Assume that $\Phi_k=(f_k,
g^{(1)}_{k},  \dots, g^{(m)}_{k})^\top$  satisfies  Eq.~\eref{LaxZuc} with $\lam=\lam_k$, where $1\leq k\leq K$. Then, the Darboux matrix $T(\lam)$ in~\eref{DarbouMatrix}  obeys Condition~\eref{DTinvariancec}, provided that  $\beta_j^{(n)}$, $\gamma_i^{(n)}$ and $\delta_{ij}^{(n)}$ ($1\leq i, j\leq m$; $0\leq n\leq N-1$) are  determined by Eqs.~\eref{UndetCoeffa},  and  the new potentials $\mb{p}'$ and $\mb{q}'$ are given by~\eref{PotentialTran}.
\end{proposition}


Seen from Propositions~\ref{P:xinvariance}--\ref{P:tinvariance}, the Darboux
matrix $T(\lam)$ makes sure that the new
eigenfunction $\Phi'= T (\lam)\,\Phi$ also
satisfies Lax representation~\eref{LaxZu} for the new potentials $\mb{p}'$ and
$\mb{q}'$ given by~\eref{PotentialTran}. That is to say, the compatibility conditions
$\Phi'_{xy}=\Phi'_{yx} $ and $\Phi'_{xt}=\Phi'_{tx} $ give the same systems~\eref{GNLS} and~\eref{GMKdV} except for $\mb{p}'$ and $\mb{q}'$ instead of $\mb{p}$ and $\mb{q}$, respectively. Therefore, we come
to the following theorem:

\begin{theorem} \label{T:Darbouxtransformation}
Suppose that $\{\Phi_k\}_{k=1}^K$ are a set of linearly-independent solutions of System~\eref{LaxZu} which correspond to a set of different spectral parameters $\{\lam_k\}_{k=1}^K$. The
eigenfunction transformation~\eref{DT} and the potential
transformations~\eref{PotentialTran} constitute the $N$-th iterated DT $ (\Phi, \mb{q}, \mb{p})
\ra (\Phi', \mb{q}', \mb{p}') $ of Systems~\eref{GNLS} and~\eref{GMKdV}, where the Darboux matrix $T(\lam)$ is determined  by Eqs.~\eref{UndetCoeffa}.
\end{theorem}

\begin{remark} \label{remarks}
With some constraints between the potentials $\mb{p}$ and $\mb{q}$,
Systems~\eref{GNLS} and~\eref{GMKdV} can be reduced to the NLEEs belonging to some known integrable  hierarchies such as the coupled NLS hierarchy~\cite{FordyVNLS} ($\mb{p} = \pm \bar{\mb{q}}$, where the bar represents complex conjugate)   and coupled modified KdV hierarchy~\cite{TsuchidamKdV} ($\mb{p} = \pm \mb{q} $). In the above way, one can also use Transformations~\eref{DT} and~\eref{PotentialTran} to constitute the $N$-th iterated DT for those reduced cases, but $\{\Phi_k\}_{k=1}^K$ are required to be particularly selected. For example, if $\mb{p} = \pm  \bar{\mb{q}}$, the functions $\{\Phi_k\}_{k=1}^K$ are chosen as~\cite{NWrVH}
\begin{subequations}
\begin{align}
&
\Phi_{(k-1)(m+1)+1}=(f_k,
g^{(1)}_{k},  \dots, g^{(m)}_{k})^\top \quad (1\leq k \leq N), \label{VNLSPhia} \\
&
\Phi_{(k-1)(m+1)+j+1} = (\pm \bar{g}^{(j)}_{k}, \overbrace{0, \dots, 0}^{j-1}, \bar{f}_k, \overbrace{0,\dots,0}^{m-j}\,)^\top \quad (1\leq k \leq N; 1\leq j\leq m), \label{VNLSPhib}
\end{align}
\end{subequations}
where $\Phi_{(k-1)(m+1)+1}$ satisfies System~\eref{LaxZu} with $\mb{p} = \pm  \bar{\mb{q}}$ and
$\lam=\lam_k$ ($1\leq k\leq N$),
and $\Phi_{(k-1)(m+1)+j+1}$ ($1\leq j \leq m$) are orthogonal to  $\Phi_{(k-1)(m+1)+1}$.
\end{remark}


\section{Multi-component Wronskian solution}

In this section, by combining the $N$-th iterated DT of Systems~\eref{GNLS} and~\eref{GMKdV} and  integrable decomposition from the KP equation~\eref{KP}  to  Systems~\eref{GNLS} and~\eref{GMKdV}, we try to construct the multi-component Wronskian solution to the KP equation~\eref{KP}.

In order to solve the functions $\beta_j^{(N-1)}$ and $\gamma_i^{(N-1)}$ ($1\leq i, j \leq m$), we rewrite Eqs.~\eref{UndetCoeffa} in the following form
\begin{eqnarray}
 & & \big( F_{K\times N}, -G^{(1)}_{K\times N},
 \cdots,  -G^{(m)}_{K\times N}
\big) \mb{X}
= \big( \lam^N_1 f_1, \cdots, \lam^N_{K} f_{K}\big)^\top, \\[1mm]
& &  \big( F_{K\times N}, -G^{(1)}_{K\times N},
\cdots, -G^{(m)}_{K\times N}
\big) \mb{Y}_i
=
\big(
\lam^N_1 g^{(i)}_{1},
\cdots,
\lam^N_{K} g^{(i)}_{K}\big)^\top  \quad (1\leq i \leq m),
\end{eqnarray}
where
\begin{eqnarray}
& & \nonumber F_{K\times N} =\left(
\begin{array}{cccc}
\vspace{1mm}  f_1 & \lam_1 f_1 & \cdots & \lam^{N-1}_1 f_1  \\
 f_2 & \lam_2 f_2 & \cdots & \lam^{N-1}_2 f_2   \\
 \vdots & \vdots & \ddots & \vdots \\
 f_{K} & \lam_{K} f_{K} & \cdots & \lam^{N-1}_{K} f_{K}
\end{array}
\right), \label{FN}
\\
& &  \nonumber  G^{(j)}_{K\times N} =\left(
\begin{array}{cccc}
\vspace{1mm}  g^{(j)}_1 & -\lam_1 g^{(j)}_1 & \cdots & (-\lam_1)^{N-1} g^{(j)}_1  \\
g^{(j)}_2 & -\lam_2 g^{(j)}_2 & \cdots & (-\lam_2)^{N-1} g^{(j)}_2  \\
 \vdots & \vdots & \ddots & \vdots \\
g^{(j)}_K & -\lam_{K} g^{(j)}_{K} & \cdots & (-\lam_{K})^{N-1} g^{(j)}_{K}
\end{array}
\right) \quad (1\leq j\leq m),
 \label{GjN}
\\[2mm]
& & \nonumber
\mb{X} = \big(\alpha^{(0)}, \alpha^{(1)}, \dots, \alpha^{(N-1)}; \beta_1^{(0)}, \beta_1^{(1)}, \dots, \beta_1^{(N-1)}; \dots; \beta_m^{(0)}, \beta_m^{(1)}, \dots, \beta_m^{(N-1)}\big)^\top,  \\
& & \nonumber
\mb{Y}_i = \big(\gamma_i^{(0)}, \gamma_i^{(1)}, \dots, \gamma_i^{(N-1)}; \delta_{i1}^{(0)}, \delta_{i1}^{(1)}, \dots, \delta_{i1}^{(N-1)}; \dots; \delta_{im}^{(0)}, \delta_{im}^{(1)}, \dots, \delta_{im}^{(N-1)}\big)^\top \quad (1\leq i\leq m).
\end{eqnarray}
By employing Cramer's rule, we can obtain the  functions $\beta_j^{(N-1)}$ and $\gamma_i^{(N-1)}$  as
\begin{eqnarray}
& & \beta_j^{(N-1)}= (-1)^{jN-1}\frac{\chi^{(1)}_j }{\tau}, \quad   \gamma_i^{(N-1)}= (-1)^{(i-1)N-1} \frac{\chi^{(2)}_i }{\tau} \quad (1\leq i, j \leq m),
\end{eqnarray}
with
\begin{subequations}
\begin{align}
& \tau = \det \big( F_{K\times N} , -G^{(1)}_{K\times N},   \cdots, -G^{(m)}_{K\times N}\big), \label{tau1}  \\
& \chi^{(1)}_j = \det \big(F_{K\times(N+1)}, -G^{(1)}_{K\times N},   \cdots, -G^{(j)}_{K\times(N-1)}, \cdots, -G^{(m)}_{K\times N}\big) \quad (1\leq j \leq m),  \label{chij1}
\\
& \chi^{(2)}_i = \det \big(F_{K\times(N-1)},  -G^{(1)}_{K\times N},   \cdots, -G^{(i)}_{K\times(N+1)}, \cdots, -G^{(m)}_{K\times N}\big) \quad (1\leq i \leq m).
\label{barchij1}
\end{align}
\end{subequations}
where $\tau$, $\chi^{(1)}_j$ and$\chi^{(2)}_i$ ($1\leq i,j \leq m$) are the multi-component Wronskians which have been proposed in Ref.~\cite{NWrVNLS1}. Thus,
the $N$-th iterated
potential transformations~\eref{PotentialTran} can be written as
\begin{align}
p'_j= p_j +2\,(-1)^{(j+1)N+1}\frac{\chi^{(1)}_j }{\tau} , \quad  q'_i=q_i-2\,(-1)^{(i-1)N-1} \frac{\chi^{(2)}_i }{\tau} \quad (1\leq i, j \leq m). \label{PotentialTran2}
\end{align}

With $\mb{p}=\mb{q}=0$, the general solution of System~\eref{LaxZu}
with $\lam= \lam_k$ ($\lam_k\neq \lam_l$; $1\leq k,l \leq K$) is given  as follows:
\begin{align}
\Phi_k =
(f_k,
g^{(1)}_{k},  \dots, g^{(m)}_{k})^\top = (a_k \me^{\frac{1}{2}\theta_k},
b^{(1)}_k \me^{-\frac{1}{2} \theta_k},  \dots,
b^{(m)}_k\me^{-\frac{1}{2}\theta_k} )^\top  \quad (1\leq k\leq K), \label{LPsolutions}
\end{align}
where
\begin{align}
\theta_k= 2 (\lam_k x - 2\,\sigma \lam_k^2  y + 4 \, \lam_k^3t ). \label{thetak}
\end{align}
It is noted that the field $u$ of Eq.~\eref{KP}  is usually assumed to be a real one. For the case $\sigma^2=1$ which corresponds to the KPII equation, we require the parameters $\lam_k$, $a_k$ and $b^{(j)}_{k}$ ($1 \leq k \leq K$; $1\leq j\leq m$) be all real constants.

\begin{proposition}
With $\Phi_k = (f_k, g^{(1)}_{k},  \dots, g^{(m)}_{k})^\top$
given by~\eref{LPsolutions} for $1 \leq k \leq K$,
the KPII equation admits the solution
$
u=2\, (\ln \tau )_{xx}  \label{MWrsolution}
$,
where the tau function $\tau$ is expressed as the following unreduced multi-component Wronskian:
\begin{align}
\tau = |A_{K\times K}\Theta_{K\times K}^+\Lambda_{K\times N}^+,  B_{K\times K}^{(1)} \Theta_{K\times K}^-\Lambda_{K\times N}^-,\dots, B_{K\times K}^{(m)} \Theta_{K\times K}^-\Lambda_{K\times N}^- |, \label{tau2}
\end{align}
with
\begin{eqnarray}
& & \nonumber A_{K\times K}=\diag\big(a_1, \dots, a_K \big),
 \quad
B_{K\times K}^{(j)} = \diag\big(-b^{(j)}_{1}, \dots, -b^{(j)}_K \big) \quad (1 \leq j \leq m),
\\[2mm]
& & \nonumber \Theta_{K\times K}^+= \diag\big(\me^{\frac{1}{2}\theta_1}, \dots, \me^{\frac{1}{2}\theta_{K}}\big), \quad
\Theta_{K\times K}^-= \diag\big(\me^{-\frac{1}{2}\theta_1}, \dots, \me^{-\frac{1}{2}\theta_{K}}\big), \\[2mm]
& & \nonumber  \Lambda_{K\times N}^+=
\begin{pmatrix}
1 & \lam_{1} & \cdots & \lam^{N-1}_{1}  \\
1 & \lam_{2} &  \cdots & \lam^{N-1}_{2}  \\
\vdots & \vdots & \ddots & \vdots \\
1 & \lam_K &\cdots & \lam^{N-1}_K
\end{pmatrix}, \quad
\Lambda_{K\times N}^-=
\begin{pmatrix}
1 & -\lam_{1} & \cdots & (-\lam_{1})^{N-1}  \\
1 & -\lam_{2} &  \cdots & (-\lam_{2})^{N-1}  \\
\vdots & \vdots & \ddots & \vdots \\
1 & -\lam_K &\cdots & (-\lam_K)^{N-1}
\end{pmatrix}.
\end{eqnarray}
\end{proposition}

\beginproof
With $\mb{p}= \mb{q}=0$, substituting Eqs.~\eref{PotentialTran2} into Eq.~\eref{KPAKNS} yields
\begin{align}
u= -2\sum_{j=1}^m p'_j q'_j= 8\sum_{j=1}^m \frac{\chi^{(1)}_j\chi^{(2)}_j}{\tau^2}.
\label{chichi}
\end{align}
On the other hand,
one can follow part (ii) of Theorem~3.3 in Ref.~\cite{NWrVNLS1} to obtain the following multi-component Wronskian identity:
\begin{align}
\sum_{j=1}^m \chi^{(1)}_j\chi^{(2)}_j = \frac{1}{4} \left( \tau\tau_{xx} -\tau^2_x \right),
\label{identity}
\end{align}
which is substituted into the right-hand side of Eq.~\eref{chichi}, giving that $u=2\,(\ln \tau)_{xx}$.  \endproof

In the case $\sigma^2 = -1$, the solution $u=2\,(\ln \tau)_{xx}$ with $\tau$ given by~\eref{tau2} also satisfies the KPI equation, but it does not meet the requirement that $u$ must be a real field.
According to Remark~\ref{remarks}, if imposing the constraint $\mb{p} = \varepsilon \bar{\mb{q}}$ ($\varepsilon = \pm 1  $) on Systems~\eref{GNLS} and~\eref{GMKdV}, one can choose
$\{\Phi_k\}_{k=1}^K$ as
\begin{subequations}
\begin{align}
&
\Phi_{(k-1)(m+1)+1}=
(a_k \me^{\frac{1}{2}\theta_k},
b^{(1)}_k \me^{-\frac{1}{2} \theta_k}, \dots,
b^{(m)}_k\me^{-\frac{1}{2}\theta_k} )^\top \quad (1\leq k \leq N), \label{Solutiona} \\
&
\Phi_{(k-1)(m+1)+j+1} = (\varepsilon \bar{b}^{(j)}_k \me^{-\frac{1}{2} \bar{\theta}_k}, \overbrace{0, \dots, 0}^{j-1}, \bar{a}_k \me^{\frac{1}{2}\bar{\theta}_k}, \overbrace{0,\dots,0}^{m-j}\,)^\top \quad (1\leq k \leq N; 1\leq j\leq m),  \label{Solutionb}
\end{align}
\end{subequations}
where the phase $\theta_k$ is the same as given in Eq.~\eref{thetak}, but the parameters $\lam_k$, $a_k$ and $b^{(j)}_{k}$ ($1 \leq k \leq N$; $1\leq j\leq m$) are all complex constants. Then, $p'_j$ and its complex conjugate $\bar{p}'_j$ are obtained as follows:
\begin{align}
p'_j=  2\,(-1)^{(j+1)N+1}\frac{\chi^{(1)}_j }{\tau} , \quad
\bar{p}'_j=  2\,\varepsilon (-1)^{(j-1)N} \frac{\chi^{(2)}_j }{\tau} \quad (1\leq j \leq m). \label{PotentialTran3}
\end{align}
where the functions $\tau$, $\chi^{(1)}_j$ and $\chi^{(2)}_j$ are given by
\begin{align}
&  \tau =
\begin{vmatrix} F_{N\times N} &  \varepsilon G^{(1)}_{N\times N}  & \cdots & \varepsilon G^{(m)}_{N\times N}
\\[0.5mm]
\bar{G}^{(1)}_{N\times N} & \bar{F}_{N\times N}  & \cdots & \mb{0}    \\
\vdots &  \vdots & \ddots & \vdots  \\
\bar{G}^{(m)}_{N\times N} & \mb{0} & \cdots  & \bar{F}_{N\times N}
\end{vmatrix}, \label{eqc01a} \\[2mm]
&
\chi^{(1)}_j =  \begin{vmatrix} F_{N\times(N+1)} & \varepsilon G^{(1)}_{N\times N} & \cdots &
\varepsilon  G^{(j)}_{N\times (N-1)}  & \cdots & \varepsilon G^{(m)}_{N\times N}
\\[0.5mm]
\bar{G}^{(1)}_{N\times(N+1)} & \bar{F}_{N\times N}  & \cdots & \mb{0} & \cdots & \mb{0} \\
\vdots &  \vdots & \ddots & \vdots & \ddots & \vdots \\
\bar{G}^{(j)}_{N\times(N+1)} & \mb{0} & \cdots  & \bar{F}_{N\times(N-1)}  & \cdots & \mb{0}  \\
\vdots & \vdots &  \ddots & \vdots &  \ddots & \vdots \\
\bar{G}^{(m)}_{N\times(N+1)} & \mb{0} & \cdots & \mb{0} & \cdots  & \bar{F}_{N\times N}   \\
\end{vmatrix} \quad (1\leq j \leq m),  \label{eqc01b} \\[1mm]
&
\chi^{(2)}_j =  \begin{vmatrix} F_{N\times(N-1)} & \varepsilon G^{(1)}_{N\times N} & \cdots &
\varepsilon G^{(j)}_{N\times (N+1)}  & \cdots & \varepsilon G^{(m)}_{N\times N}
\\[0.5mm]
\bar{G}^{(1)}_{N\times(N-1)} & \bar{F}_{N\times N}  & \cdots & \mb{0} & \cdots & \mb{0} \\
\vdots &  \vdots & \ddots & \vdots & \ddots & \vdots \\
\bar{G}^{(j)}_{N\times(N-1)} & \mb{0} & \cdots  & \bar{F}_{N\times(N+1)}  & \cdots & \mb{0}  \\
\vdots & \vdots &  \ddots & \vdots &  \ddots & \vdots \\
\bar{G}^{(m)}_{N\times(N-1)} & \mb{0} & \cdots & \mb{0} & \cdots  & \bar{F}_{N\times N}   \\
\end{vmatrix} \quad (1\leq j \leq m),  \label{eqc01c}
\end{align}
where $ F_{N\times M } = A_{N\times N} \Theta^+_{N\times N}  \Lambda^+_{N \times M} $, $
G^{(j)}_{N\times M } = B^{(j)}_{N\times N} \Theta^-_{N\times N}  \Lambda^-_{N\times M} $ ($M=N-1, N, N+1$). Via the multi-component Wronskian identity~\eref{identity},
we know that
\begin{align}
u= -2 \,\varepsilon \sum_{j=1}^m |p'_j|^2 = 8\sum_{n=1}^m \frac{\chi^{(1)}_j\chi^{(2)}_j}{\tau^2}
=\frac{2(\tau\tau_{xx} - \tau^2_x)}{\tau^2}=2\,(\ln \tau)_{xx} ,
\end{align}
which implies that $u$ is a real function.

\begin{proposition}
With $\Phi_k = (f_k, g^{(1)}_{k},  \dots, g^{(m)}_{k})^\top$
given by Eqs.~\eref{Solutiona} and~\eref{Solutionb} for $1 \leq k \leq K$, the KPI equation admits the  solution $ u=2\,(\ln \tau )_{xx} $, where the tau function $\tau $ is a reduced multi-component Wronskian given in Eq.~\eref{eqc01a}.
\end{proposition}


\section{Soliton solutions to the KP  equation}

In this section, we  explore the soliton solutions to the KPII equation generated by the unreduced multi-component Wronskian~\eref{tau2}, and to the KPI equation generated by the reduced multi-component Wronskian~\eref{eqc01a}.

\subsection{Fully-resonant soliton solutions to the KPII equation}

As for the KPII equation, one might naturally ask whether the soliton solutions generated by the multi-component Wronskian~\eref{tau2} are different from those by the single Wronskian~\eref{KPSoliton}. Since the multi-component Wronskian~\eref{tau2} with $m>1$ generates singular solutions for the generic choice of parameters, we  in this subsection discuss the soliton solutions in $u =2\, (\ln\tau)_{xx} $ with
\begin{align}
\tau = |A_{2N\times 2N}\Theta_{2N\times 2N}^+\Lambda_{2N \times N}^+,  B_{2N\times 2N}^{(1)} \Theta_{2N\times 2N}^-\Lambda_{2N\times N}^-|.  \label{m1tau}
\end{align}
For convenience, we  take the coefficient matrix
$B_{2N\times 2N}^{(1)} = B_{2N\times 2N} = \diag(-b_{1}, \dots, -b_{2N} ) $ with $ b_k  =b^{(1)}_{k} $ for $k\in [2N]:=\{1,2,\dots, 2N\}$, and  for comparison with the soliton solutions in Refs.~\cite{Biondini2006,Biondini2007PRL,Chak2008,Chak2009} we  assume that
$ \theta_k=  - \kappa_k x -\sigma \kappa_k^2 y - \kappa_k^3 t
$,  where $\kappa_k =-2\,\lam_k$ and $\{\kappa_k \}_{k=1}^{2N}$ are without loss of generality ordered as  $\kappa_1 < \kappa_2 \dots < \kappa_{2N} $.

By using the Laplace expansion technique and Binet-Cauchy theorem, we can expand the  function $\tau$ in Eq.~\eref{m1tau} as follows:
\begin{align}
\tau =
\sum_{\substack {\mathcal{I}_N\cap\mathcal{J}_N=\varnothing,
\\\mathcal{I}_N\cup\mathcal{J}_N=[2N]}}
(-1)^{N+\sum\limits_{n=1}^N i_n} \prod\limits_{n=1}^N a_{i_n} b_{j_n}\!\!\!
\prod_{1\leq n < l \leq N}
\Big(\frac{1}{2} \kappa_{i_n} -\frac{1}{2} \kappa_{i_l}\Big)\Big(\frac{1}{2} \kappa_{j_n}-\frac{1}{2} \kappa_{j_l}\Big)
\exp \bigg[ \frac{1}{2}\sum_{n=1}^{N}(\theta_{i_n} - \theta_{j_n})\bigg],  \label{tauexpansion}
\end{align}
where  $\mathcal{I}_N = \{i_1, \dots, i_N \}$ and $\mathcal{J}_N = \{j_1, \dots, j_N \}$
are two subsets of $[2N]$ with $1\leq i_1 < \dots < i_N \leq 2N$ and
$1\leq j_1 < \dots < j_N\leq 2N$. In order to ensure that the solution $u$ to the KPII equation resulting from the function $\tau$ in Eq.~\eref{m1tau} is non-trivial and non-singular,  the coefficients $a_1, \dots, a_{2N}, b_1, \dots, b_{2N}$ are required to satisfy the following three conditions:
\begin{enumerate}
\item[(i)] $a_k $ and $b_k$  are not equal to zero for the same $k\in [2N]$;

\item[(ii)] $|\{a_k|a_k \neq 0  \}| = N + L $ and $|\{b_k|b_k \neq 0  \}|
= N + M  $, where $1 \leq L,M \leq N$ and $|\cdot|$ denotes the number of elements in a set;

\item[(iii)] $a_k a_{k+1} b_k b_{k+1} \leq 0$ for $1\leq k \leq 2N-1$.
\end{enumerate}
Here, condition (i) requires that the function $\tau$  does not reduce to zero;
condition (ii) requires that the function $\tau$  contains at least two exponentials;
condition (iii) requires that the function $\tau$ is sign definite, so that  $\tau$ has no zeros  for all $(x,y,t)\in \mathbb{R}^3$.

Considering that there might be some zeros  among the coefficients
$a_1, \dots, a_{2N}, b_1, \dots, b_{2N}$, we assume that $ a_{i^*_n } =0 $ for $n \in [L']$ and $ b_{j^*_n } =0 $ for $n \in [M']$, where $L' = N-L $,  $M' = N-M $,
$1\leq i^*_1 < \dots < i^*_{L'} \leq 2N$, $1\leq j^*_1 < \dots < j^*_{M'} \leq 2N$ and $i^*_n \neq j^*_l $.
Thus, the function $\tau $ in Eq.~\eref{tauexpansion} can be expressed as
\begin{align}
\nonumber \tau  = & \sum_{\substack {\mathcal{I}_M\cap\mathcal{J}_L=\varnothing,
\mathcal{I}_M\cap\mathcal{J}^*_{M'}=\varnothing,  \mathcal{J}_L\cap\mathcal{I}^*_{L'}=\varnothing
\\\mathcal{I}_M\cup\mathcal{J}_L=[2N]\setminus (\mathcal{I}^*_{L'}\cup \mathcal{J}^*_{M'}) }} (-1)^{N +\sum\limits_{n=1}^{M} i_n
+ \sum\limits_{n=1}^{M'} j^*_n }
\prod_{n=1}^{L'} \Upsilon_{i^*_n}
\prod_{n=1}^{M'}  \Upsilon_{j^*_n}
\prod\limits_{n=1}^{M} a_{i_n}
\prod\limits_{n=1}^{L}
b_{j_n}   \\
&
\times \prod_{1\leq n < l \leq M}
\Big(\frac{1}{2} \kappa_{i_n} -\frac{1}{2} \kappa_{i_l}\Big)\!\! \prod_{1\leq n < l \leq L} \Big(\frac{1}{2} \kappa_{j_n} -\frac{1}{2} \kappa_{j_l} \Big)
\exp \bigg[ \frac{1}{2}\bigg(\sum_{n=1}^{M}\theta_{i_n} + \sum_{n=1}^{M'}\theta_{j^*_n}
- \sum_{n=1}^{L}\theta_{j_n} -
\sum_{n=1}^{L'}\theta_{i^*_n} \bigg) \bigg], \label{tauexpansion1}
\end{align}
where $\mathcal{I}_M = \{i_1, \dots, i_M \}$, $\mathcal{J}_L = \{j_1, \dots, j_L \}$,
$\mathcal{I}^*_{L'} = \{i^*_1, \dots, i^*_{L'} \}$, $\mathcal{J}^*_{M'} = \{j^*_1, \dots, j^*_{M'} \}$, $\Upsilon_{i^*_n}$ and $\Upsilon_{j^*_n}$ are given by
\begin{align}
 & \nonumber  \Upsilon_{i^*_n} = b_{i^*_n}
\prod_{j_l < i^*_n } \Big(\frac{1}{2} \kappa_{j_l} - \frac{1}{2} \kappa_{i^*_n}\Big)\prod_{j_l> i^*_n}
\Big(\frac{1}{2} \kappa_{i^*_n}-\frac{1}{2} \kappa_{j_l} \Big) \quad (1\leq  n\leq L'),\\
& \nonumber \Upsilon_{j^*_n} = a_{j^* _n} \prod_{i_l < j^*_n }
\Big(\frac{1}{2} \kappa_{i_l}-\frac{1}{2} \kappa_{j^*_n} \Big)\prod_{i_l> j^*_n} \Big(\frac{1}{2} \kappa_{j^*_n} -\frac{1}{2} \kappa_{i_l}\Big) \quad
(1 \leq n \leq M').
\end{align}

With scaling transformation,
the function $\tau$ in Eq.~\eref{tauexpansion1} is further equivalent to
\begin{align}
\nonumber  \tau' & = \tau \big/\bigg\{
 \sum_{\substack {\mathcal{I}_M\cap\mathcal{J}_L=\varnothing,
\mathcal{I}_M\cap\mathcal{J}^*_{M'}=\varnothing,  \mathcal{J}_L\cap\mathcal{I}^*_{L'}=\varnothing
\\\mathcal{I}_M\cup\mathcal{J}_L=[2N]\setminus (\mathcal{I}^*_{L'}\cup \mathcal{J}^*_{M'}) }}
(-1)^{L' + \sum\limits_{n=1}^{M'} j^*_n }
\prod_{n=1}^{L'} \Upsilon_{i^*_n}
\prod_{n=1}^{M'}  \Upsilon_{j^*_n}
\prod\limits_{n=1}^{M} a_{i_n}
\prod\limits_{n=1}^{L}
a_{j_n}   \\
 & \hspace{5mm} \times  \Big(\frac{1}{2} \Big)^{\frac{M(M+1)}{2} + \frac{L(L+1)}{2}}
\exp \bigg[ \frac{1}{2}\bigg(\sum_{n=1}^{M'}\theta_{j^*_n} - \sum_{n=1}^{M}\theta_{i_n}
-\sum_{n=1}^{L}\theta_{j_n} -
\sum_{n=1}^{L'}\theta_{i^*_n}  \bigg) \bigg]
\bigg\}
\\
\nonumber & =
\sum_{\substack {\mathcal{I}_M\cap\mathcal{J}_L=\varnothing,
\mathcal{I}_M\cap\mathcal{J}^*_{M'}=\varnothing,  \mathcal{J}_L\cap\mathcal{I}^*_{L'}=\varnothing
\\\mathcal{I}_M\cup\mathcal{J}_L=[2N]\setminus (\mathcal{I}^*_{L'}\cup \mathcal{J}^*_{M'}) }} (-1)^{L +\sum\limits_{n=1}^{M} i_n} \prod\limits_{n=1}^{L}  b'_{j_n}
\\
& \hspace{5mm} \times
\prod_{1\leq n < l \leq M}
(\kappa_{i_n}- \kappa_{i_l} )
\prod_{1\leq n < l \leq L}( \kappa_{j_n} - \kappa_{j_l})
\exp \bigg(\sum_{n=1}^{M}\theta_{i_n}  \bigg) \quad ( b'_{j_n} = b_{j_n}/ a_{j_n}), \label{tauexpansion2}
\end{align}
in generating the solution $u$ of the KPII equation. Without loss of generality, we suppose that $\mathcal{I}_M\cup \mathcal{J}_L = [M+L] $ and
$\mathcal{I}^*_{L'}\cup \mathcal{J}^*_{M'} = \{M+L+1, \dots, 2N \}$. We notice that the
function $\tau'$ in Eq.~\eref{tauexpansion2} contains all possible combinations of $M$
phases out of $\{\theta_1,  \dots, \theta_{M+L}\}$, which is the same as the case of
a family of fully-resonant soliton solutions to the KPII equation discussed in Ref.~\cite{Biondini2003}. Therefore, we immediately come to the following results:
\begin{proposition}
The dominant exponentials of the function $\tau'$ in adjacent regions of the $xy$-plane as $\sigma y \ra \pm \infty$ contain $M-1$ common phases and differ by only one phase, and the transition between two such  exponentials occurs along the line defined by $L_{ij}: \theta_i =\theta_j $ where $j =  i+M  $ for $i\in [L]$ as $\sigma y \ra  \infty$ and $j = i + L$ for $i\in [M]$ as $\sigma y \ra  -\infty$. In a neighborhood of the transition line $L_{ij}$ as $\sigma y \ra \pm \infty$, the asymptotic behavior of $u = 2\,(\ln \tau')_{xx}$ is determined by
\begin{align}
u \sim u^\pm_{[i,j]} = 2\,\frac{\partial^2}{\partial x^2 } \ln\big(C^\pm_i \me^{\theta_i} +C^\pm_j \me^{\theta_j} \big)  =  A_{[i,j]}\sech^2\bigg(\mb{K}_{[i,j]} \cdot \mb{r}  + \Omega_{[i,j]} t + \ln \frac{C^\pm_j}{C^\pm_i} \bigg), \label{AsypSoliton}
\end{align}
with
\begin{align}
& \nonumber C^+_i= (-1)^i b'_j\prod\limits_{i+1 \leq  n \leq j-1}
(\kappa_i - \kappa_n)\prod_{1\leq  n \leq i-1}(\kappa_n-\kappa_j)
\prod_{j+1 \leq n  \leq M+L}(\kappa_j  - \kappa_n ), \\
& \nonumber C^+_j = (-1)^j b'_i\prod\limits_{i+1 \leq  n \leq j-1}
(\kappa_n - \kappa_j )\prod_{1\leq  n \leq i-1}(\kappa_n -\kappa_i )
\prod_{j+1 \leq n  \leq M+L}(\kappa_i - \kappa_n ), \\
& \nonumber C^-_i= (-1)^i b'_j \prod_{1\leq  n \leq i-1}(\kappa_n-\kappa_i)
\prod_{j+1 \leq n  \leq M+L}(\kappa_i  - \kappa_n )
\prod\limits_{i+1 \leq  n \leq j-1}
( \kappa_n - \kappa_j), \\
& \nonumber C^-_j = (-1)^j b'_i \prod_{1\leq  n \leq i-1}(\kappa_n-\kappa_j)
\prod_{j+1 \leq n  \leq M+L}(\kappa_j  - \kappa_n )
\prod\limits_{i+1 \leq  n \leq j-1}
(\kappa_i - \kappa_n ), \\
& \nonumber \mb{r} =(x,y), \quad   \mb{K}_{[i,j]}  = \left[ \frac{1}{2}(\kappa_j-\kappa_i), \frac{1}{2\,\sigma}(\kappa^2_j-\kappa^2_i) \right],  \\
 & \nonumber A_{[i,j]}  =  \frac{1}{2}(\kappa_j-\kappa_i)^2,  \quad
\Omega_{[i,j]}  = \frac{1}{2} (\kappa^3_j-\kappa^3_i),
\end{align}
where $u^\pm_{[i,j]}$ defines the asymptotic line soliton $[i,j]$ as $\sigma y \ra \pm \infty$, and $A_{[i,j]}$,  $\mb{K}_{[i,j]}$ and  $\Omega_{[i,j]} $ respectively represent the amplitude, wave vector and frequency of asymptotic soliton.
\end{proposition}

\begin{figure}[H]
\begin{minipage}[t]{0.48\linewidth} 
\centering
\includegraphics[scale=0.65]{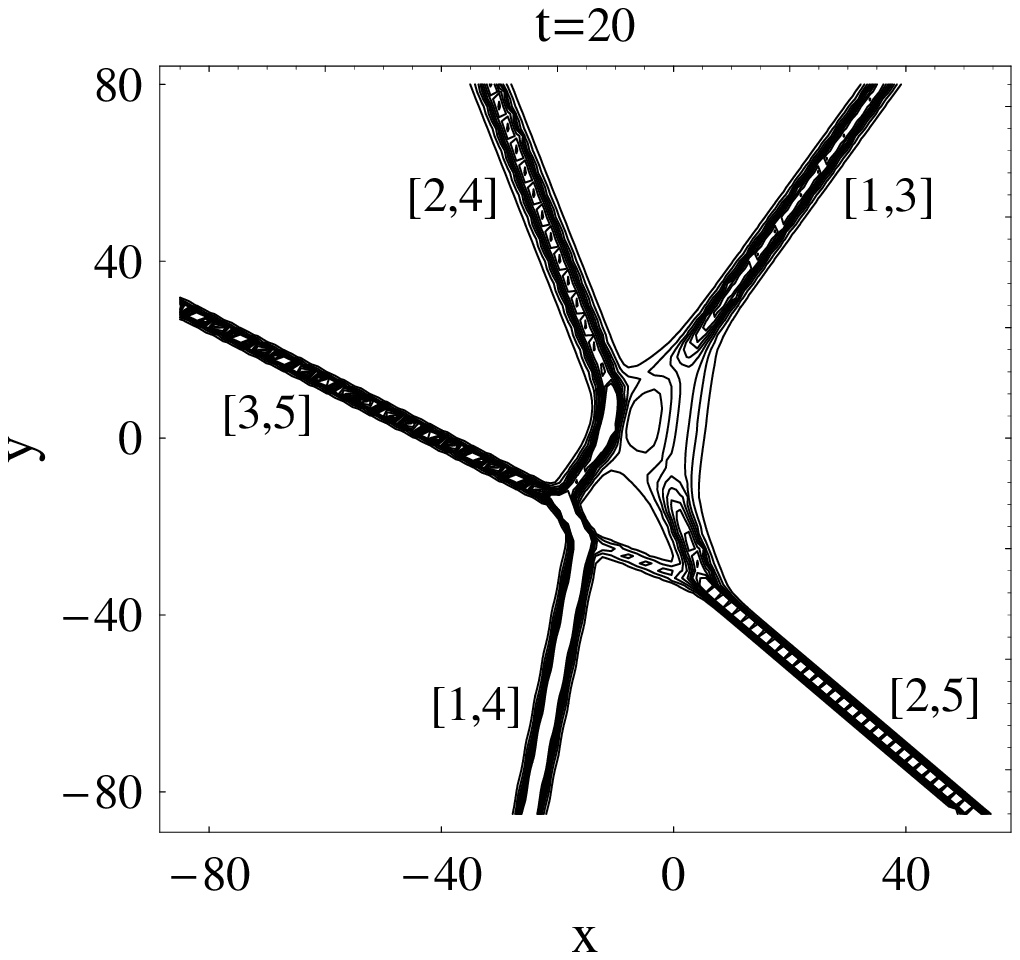}
 \caption{ \linespread{1.3} \selectfont A fully-resonant $(3,2)$-soliton solution to the KPII equation generated by
 the  function $\tau'$ in Eq.~\eref{tauexpansion2} with the parameters chosen as $L=3$, $M=2$, $\sigma=1$,  $b'_1=b'_2=b'_3=b'_4=b'_5=1$, $\kappa_1= -0.8$, $\kappa_2= -0.35$, $\kappa_3= 0.25$, $\kappa_4= 0.65$ and $\kappa_5= 1.25$.
  \label{23soliton} }
\end{minipage}\hspace{5mm}
\begin{minipage}[t]{0.48\linewidth}
\centering
\includegraphics[scale=0.65]{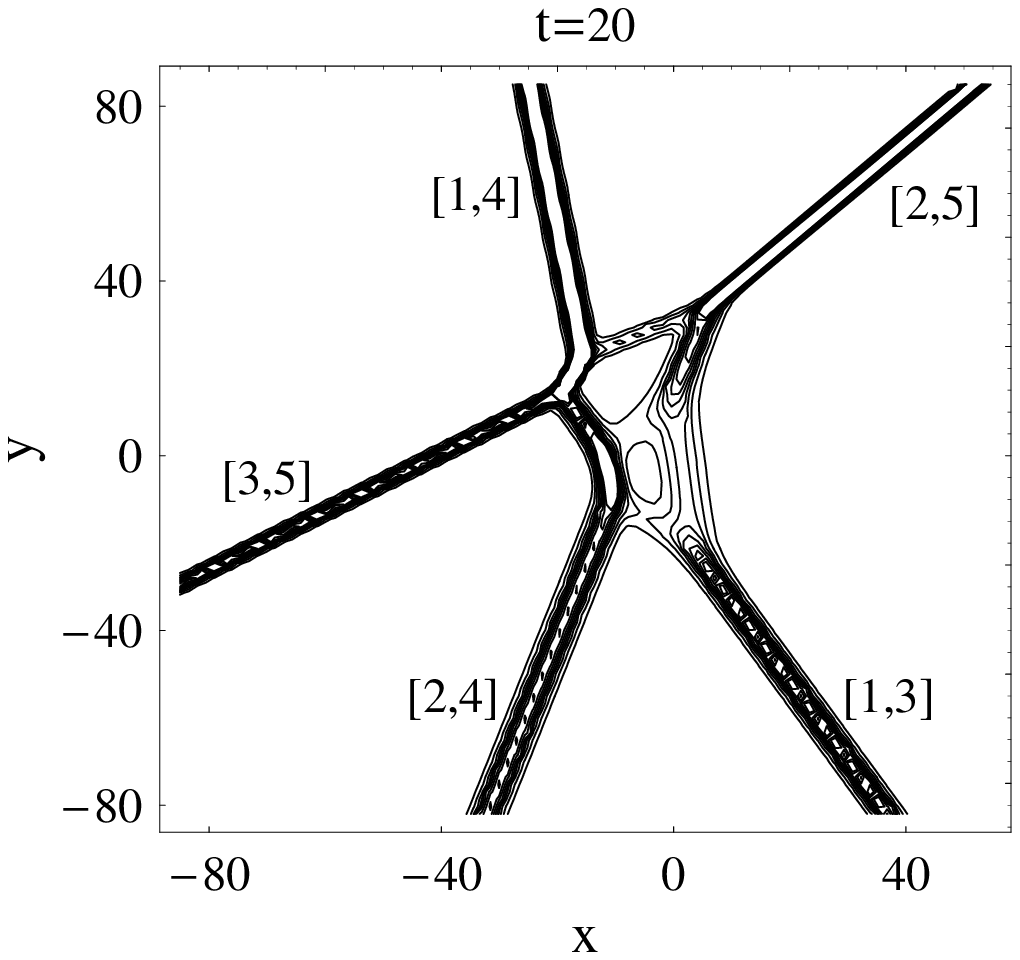}
\caption{ \linespread{1.3} \selectfont A fully-resonant $(2,3)$-soliton solution to the KPII equation generated by
 the  function $\tau'$ in Eq.~\eref{tauexpansion2} with the same parameters as those in Figure~\ref{23soliton} except for $\sigma=-1$.
 \label{32soliton} }
\end{minipage}
\end{figure}

We remark that the function $\tau'$ in Eq.~\eref{tauexpansion2} generates two families of fully-resonant soliton solutions: one is $(L,M)$-soliton configuration with  $L$ asymptotic solitons as $y\ra \infty $ and $M$ asymptotic solitons as $y\ra -\infty $ for $\sigma=1$ (see Figure~\ref{23soliton}), and the other is $(M,L)$-soliton configuration with  $M$ asymptotic solitons as $y\ra \infty $ and $L$ asymptotic solitons as $y\ra -\infty $ for $\sigma=-1$ (see Figure~\ref{32soliton}).  Such two families of solutions correspond to two special cases of the soliton solutions generated by the function $\tau $ in Eq.~\eref{KPSoliton} because the soliton amplitude $A_{[i,j]}$, wave vector $\mb{K}_{[i,j]}$ and frequency $\Omega_{[i,j]} $ in the asymptotic expression~\eref{AsypSoliton} are the same as those given in
Refs.~\cite{Biondini2006,Chak2008,Chak2009}.


\subsection{Ordinary $N$-soliton solution to the KPI equation}

In this subsection, we  are devoted to discussing the soliton solutions to the KPI equation generated by the function $\tau$ in Eq.~\eref{eqc01a} with  $m=1$ and $\varepsilon =-1$, i.e.,
\begin{align}
\tau =
\begin{vmatrix} F_{N\times N} & -G^{(1)}_{N\times N}
\\
\bar{G}^{(1)}_{N\times N} & \bar{F}_{N\times N}
\end{vmatrix}, \label{TwoWr}
\end{align}
where $\tau$ is a real-valued function and has no zeros for all $(x,y, t)\in \mathbb{R}^3$.
Our consideration is based on the following two facts: (a) The solutions generated by the $(m+1)$-component Wronskian~\eref{eqc01a} do not contain  more valid free parameters than those by the two-component Wronskian~\eref{TwoWr};
(b) The function $\tau$ in Eq.~\eref{eqc01a} with $\varepsilon =1$ has zeros for some $(x,y,t) \in \mathbb{R}^3$. For convenience,  we define that $ b_k  =b^{(1)}_{k} $ and $\lam_k = \frac{1}{4} \left(\mu_k + \sigma \nu _k\right)$ for $ k\in [N]$, where $\sigma = \pm \mi$, $\mu _k$ and $\nu _k$ are real constants. Without loss of generality, we assume that $a_k =1, b_k \neq 0$ for $ k\in [N]$ and
$\{\nu_k\}_{k=1}^N$ are well ordered as $\nu_1 < \nu_2 < \dots <
\nu_N$.

\begin{figure}[H]
\begin{minipage}[t]{0.48\linewidth} 
\centering
\includegraphics[scale=0.65]{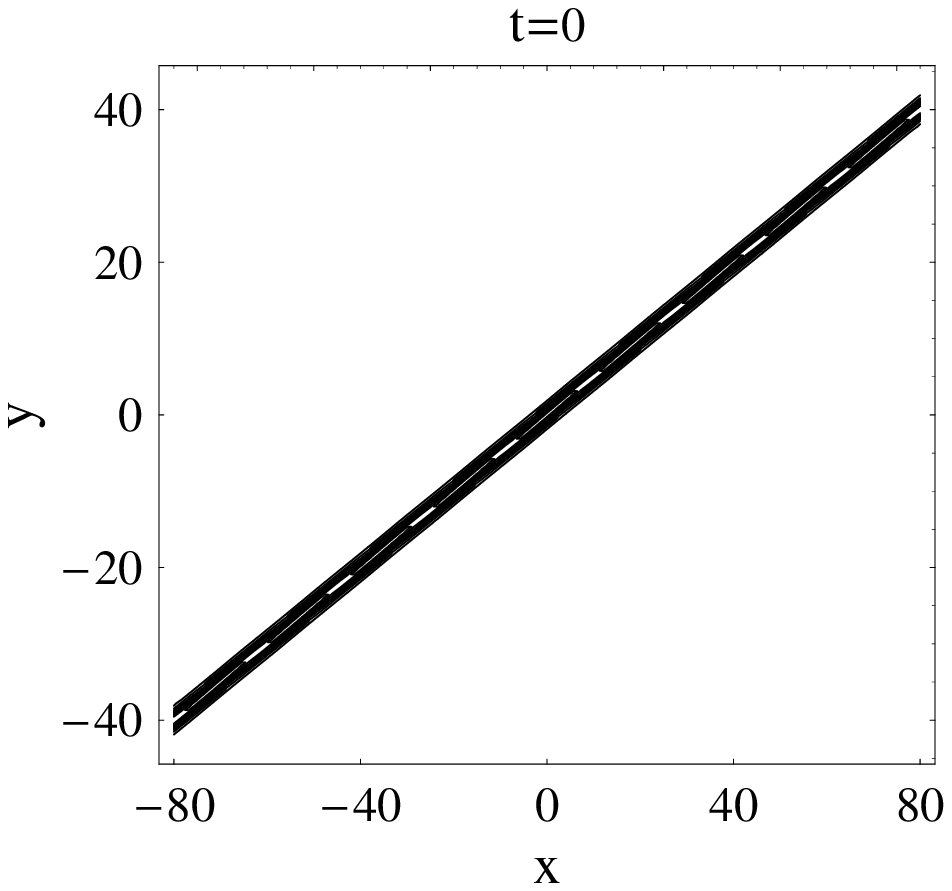}
 \caption{ \linespread{1.3} \selectfont A one-soliton solution to the KPI equation generated by
 the  function $\tau$ in Eq.~\eref{TwoWr} with the parameters chosen as $N=1$, $\sigma=\mi$, $\mu_1=1$, $\nu_1=2$ and $b_1=1$.
  \label{Ionesoliton} }
\end{minipage}\hspace{5mm}
\begin{minipage}[t]{0.48\linewidth}
\centering
\includegraphics[scale=0.65]{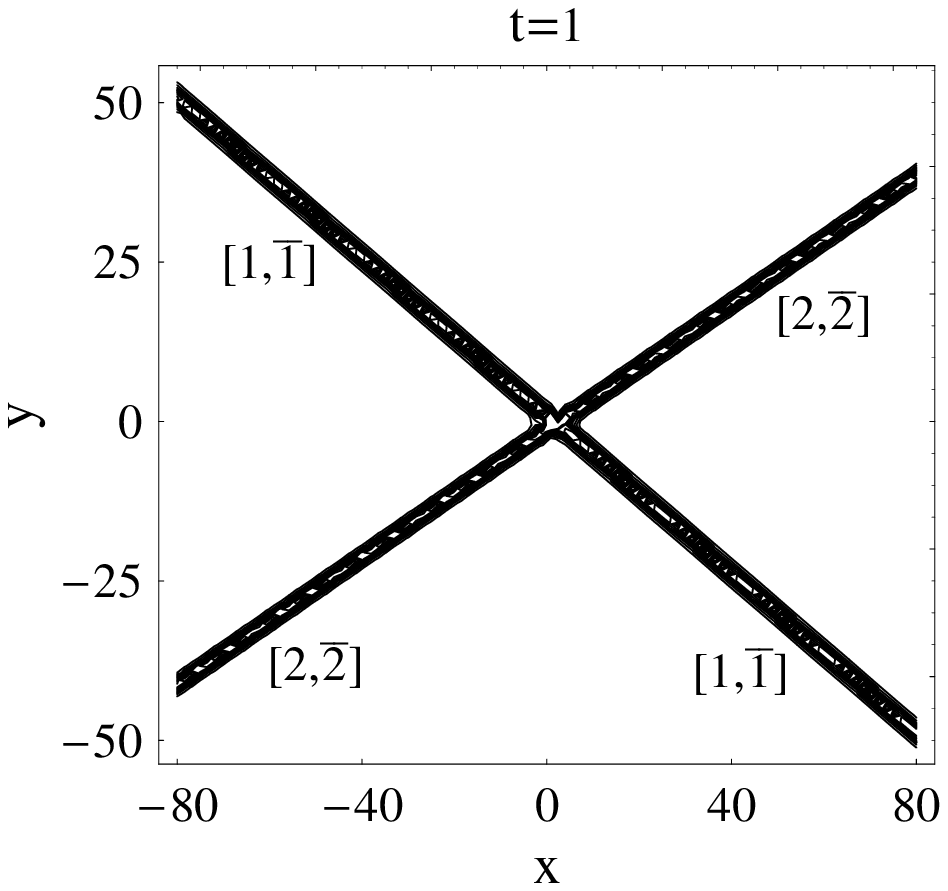}
\caption{ \linespread{1.3} \selectfont An obliquely-colliding two-soliton solution
 to the KPI equation generated by
 the  function $\tau$ in Eq.~\eref{TwoWr}  with the parameters chosen as $N=2$, $\sigma=\mi$,  $\mu_1=\mu_2=1$, $\nu_1=-1.6$, $\nu_2=2$ and $b_1=b_2=1$.
 \label{Itwosoliton} }
\end{minipage}
\end{figure}

For the case $N=1$, the solution $u=2(\ln \tau)_{xx}$ with $\tau $ given by Eq.~\eref{TwoWr} can be expressed as
\begin{align}
u = \frac{1}{2}\,\mu_1^2 \text{sech}^2  \Big[\frac{1}{2} \mu_1 x - \frac{1}{2}\mu _1 \nu _1 y + \frac{1}{8}\, \mu_1\! \left(\mu _1^2-3 \nu _1^2\right) t -  \ln|b_1| \Big] ,
\end{align}
which describes the one-soliton solution to the KPI equation (see Figure~\ref{Ionesoliton}),  where the amplitude, wave vector and frequency of soliton are respectively given by  $\frac{1}{2}\,\mu_1^2$, $(\frac{1}{2}\,\mu_1, - \frac{1}{2}\,\mu _1 \nu _1)$ and $\frac{1}{8}\, \mu _1\! \left(\mu _1^2-3 \nu _1^2\right) $. It is mentioned that such line-soliton solution is in accordance with those obtained by the Hirota method~\cite{Satsuma1976} and inverse scattering transform~\cite{Infeld1990}.

For the case $N \ge 2$, we can employ the results of Ref.~\cite{NWrVNLS1} to analyze the asymptotic behavior of the function $\tau $ in Eq.~\eref{TwoWr}.
According to Lemma 4.1 in Ref.~\cite{NWrVNLS1}, we know that the function $\tau$ in Eq.~\eref{TwoWr}
 can be expanded as the sum of such exponential terms as $ \exp \big[\frac{1}{2}\sum_{k=1}^N\! \left(r_k \theta_k +s_k \bar{\theta}_k\right) \big]$, where $r_k, s_k \in\{-1,1\}$,
$\big|\{r_k|r_k=-1\} \big|=\big|\{s_k|s_k= -1 \} \big| $ and
$\big|\{r_k|r_k=1\} \big|=\big|\{s_k|s_k= 1 \} \big|$.
Further  using Theorem 4.4 in Ref.~\cite{NWrVNLS1}, we obtain that
along the line $\theta_n + \bar\theta_n =0$ ($1\leq n \leq N$) as $y\ra \pm \infty$ for finite values of $t$, the solution $u=2(\ln \tau)_{xx}$ is asymptotically determined by
\begin{align}
\nonumber u \sim u^{\pm}_{[n, \bar{n}]} & =  2\,\frac{\partial^2}{\partial x^2 } \ln
\Big( \zeta^\pm_n \exp \Big\{  \frac{1}{2}\big[ \theta_n + \bar{\theta}_n
+ \underset{k\neq n}{\sum} \epsilon^{\pm}_{kn}(\theta_k +
\bar{\theta}_k)\big] \Big \}   \\
\nonumber & \hspace{2cm} + \eta^\pm_n \exp \Big\{\frac{1}{2}\big[-\theta_n   -\bar{\theta}_n + \underset{k\neq n}{\sum} \epsilon^{\pm}_{kn}(\theta_k +
\bar{\theta}_k)\big] \Big \} \Big)  \\
& = \frac{1}{2}\,\mu_n^2 \,\sech^2\!\Big(\frac{\theta_n + \bar{\theta}_n}{2}
+ \ln \sqrt{\zeta^\pm_{n}/\eta^\pm_{n} }\, \Big)\quad  (1\leq n \leq N),
\label{Asymptoticexpressions}
\end{align}
where $u^\pm_{[n, \bar{n}]}$ defines the $n$-th asymptotic line soliton $[n,\bar{n}]$ as $y \ra \pm \infty$, $\zeta_{n}^\pm$ and $\eta_{n}^\pm$ correspond to the coefficients of two dominant exponentials  in the  expansion of $\tau$, the parameters
$\epsilon^\pm_{kn}$'s are defined as
\begin{eqnarray}
& & \epsilon^-_{kn} =  \left\{
\begin{array}l
 -1, \quad  \text{for $k\in \mathcal{B}_n^{(\rm{I})} \cup \mathcal{B}_n^{(\rm{I\!I})}$ }, \\[0.5mm]
\ 1, \quad   \ \  \text{for $k\in \mathcal{B}_n^{(\rm{I\!I\!I})} \cup \mathcal{B}_n^{(\rm{I\!V})} $},
\end{array}
\right.  \quad \epsilon^+_{kn}  =  \left\{
\begin{array}l
\  1, \quad \ \  \text{for $k\in \mathcal{B}_n^{(\rm{I})} \cup \mathcal{B}_n^{(\rm{I\!I})}$}, \\[0.5mm]
-1, \quad  \text{for $k\in \mathcal{B}_n^{(\rm{I\!I\!I})} \cup \mathcal{B}_n^{(\rm{I\!V})}$},
\end{array}
\right.
\end{eqnarray}
with $\mathcal{B}_n^{(\rm{I})} =\{l| \mu_l>0, l=1,\dots,n-1\}$,
$\mathcal{B}_n^{(\rm{I\!I})} =\{l|\mu_l<0, l=n+1,\dots,N\}$,
$\mathcal{B}_n^{(\rm{I\!I\!I})} =\{l| \mu_l<0, l=1,\dots,n-1\}$
and $\mathcal{B}_n^{(\rm{I\!V})} =\{l|\mu_l>0, l=n+1,\dots,N\}$.

\begin{figure}[H]
\begin{minipage}[t]{0.48\linewidth} 
\centering
\includegraphics[scale=0.65]{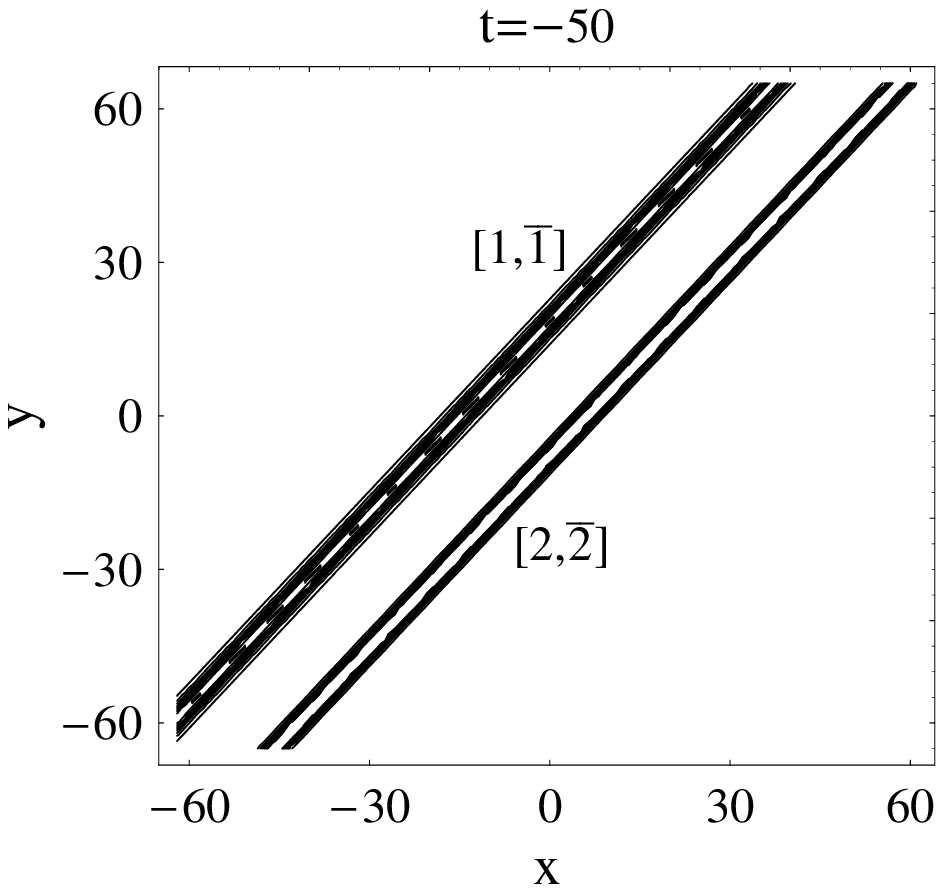}
 \caption{ \linespread{1.3} \selectfont Parallel propagation of two line solitons  generated by
 the  function $\tau$ in Eq.~\eref{TwoWr} with the parameters chosen as $N=2$, $\sigma=\mi$,  $\mu_1=1$, $\mu_2=1.5$, $\nu_1=\nu_2=0.8$ and $b_1=b_2=1$.
  \label{Para2soliton} }
\end{minipage}\hspace{5mm}
\begin{minipage}[t]{0.48\linewidth}
\centering
\includegraphics[scale=0.65]{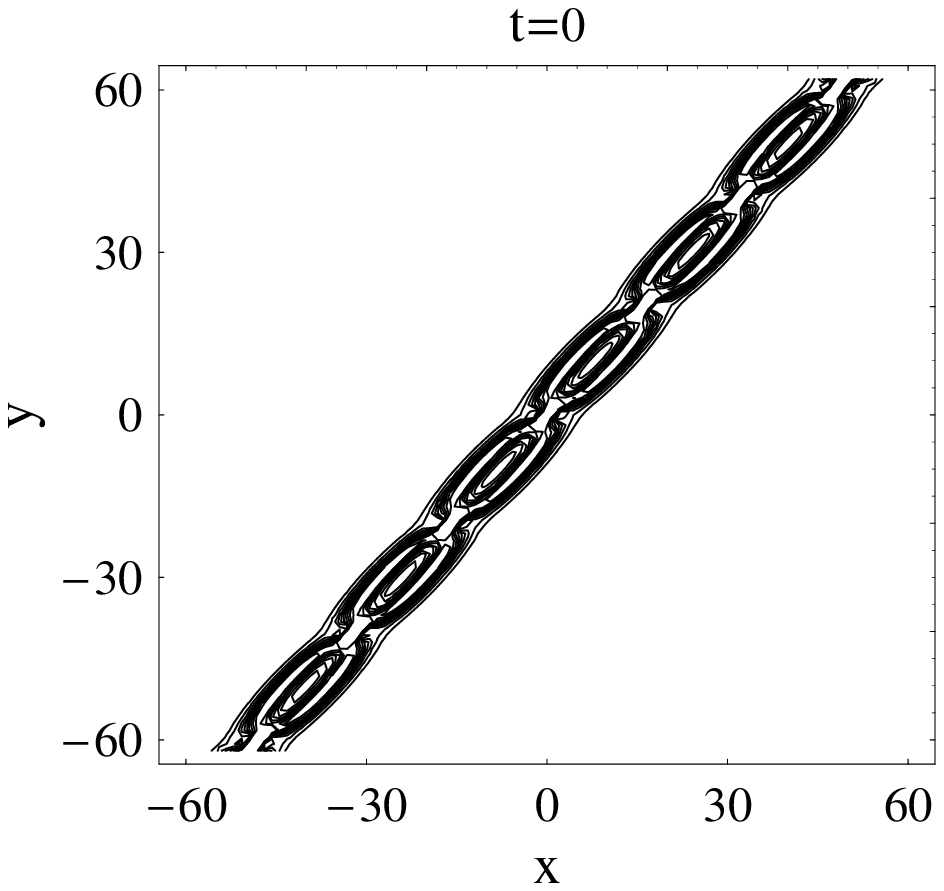}
\caption{ \linespread{1.3} \selectfont Bound state between two line solitons  at the moment of collision generated by the  function $\tau$ in Eq.~\eref{TwoWr}  with the same parameters  as those in Figure~\ref{Para2soliton}.
 \label{Bound2soliton} }
\end{minipage}
\end{figure}

The asymptotic expression~\eref{Asymptoticexpressions} shows that the $n$-th asymptotic soliton as $y\ra \infty$ has the same velocity and amplitude as the $n$-th asymptotic soliton as $y\ra -\infty$ except for a slight phase shift, as displayed in Figure~\ref{Itwosoliton}. Therefore, the $N$-soliton solution generated by the function $\tau$ in Eq.~\eref{TwoWr}  reflect only the ordinary elastic collisions of line solitons in the KPI equation. Particularly with $\nu_n =\nu_k $ ($n\neq k$), the $n$-th and $k$-th asymptotic solitons propagate  parallel to each other (see Figure~\ref{Para2soliton}), but they exhibit the bound state~\cite{Boundsoliton} at the moment of collision (see Figure~\ref{Bound2soliton}).
It is an interesting question whether the KPI equation also admits the inelastic  soliton collisions (in which the numbers, amplitudes or directions of asymptotic solitons are not the same as $y\ra \pm \infty$), like the cases in the KPII equation.

\section{Conclusions}

In this paper, we have  constructed the  $N$-th iterated DT for the second- and third-order $m$-coupled AKNS systems~\eref{GNLS} and~\eref{GMKdV}, and rigorously proved the form-invariance of their Lax representation~\eref{LaxZu}. Then, using Cramer's rule and based on the integrable decomposition from the KP equation~\eref{KP} to Systems~\eref{GNLS} and~\eref{GMKdV}, we have derived the unreduced multi-component Wronskian solution to the KPII equation and a reduced multi-component Wronskian solution to the KPI equation, which implies that the multi-component Wronskian is a third form of the tau function of the KP equation in addition to the Wronskian and Grammian. Further, by analyzing algebraic properties of the unreduced and reduced two-component Wronskians, we have obtained
two families of fully-resonant line-soliton solutions which, in general, contain unequal numbers of asymptotic solitons as $y \ra \mp \infty$ to the KPII equation, and the ordinary $N$-soliton solution which generally describes the elastic  collisions of line solitons to the KPI equation. In particular, we have found that the KPI line solitons which propagate in parallel can exhibit the bound state at the moment of collision. It is remarkable that the double Wronskian representation of the $N$-soliton solution  to the KPI equation has been revealed in this work for the first time.

\section*{Acknowledgements}

TX would like to thank the helpful discussions with Professor
G. Biondini.  This work has been supported by the Science Foundation of  China University of Petroleum, Beijing (No. BJ-2011-04), and  by the National Natural Science Foundations of China under Grants No. 11071257 and No. 11101421.

\newpage
\begin{appendix}

\section{Proof of Lemma~\ref{L:uproperties}}
\renewcommand{\theequation}{A.\arabic{equation}}
\setcounter{equation}{0} \label{appendixA}

\beginproof First, we make use of Eqs.~\eref{UndetCoeffa} to compute $T^*(\lam_k)$ as
\begin{align}
T^*(\lam_k)=
\begin{pmatrix}
\Delta_0(\lam_k) &   -\Delta_1(\lam_k) & \cdots & (-1)^m\Delta_m(\lam_k) \\
\varpi^{(1)}_k\Delta_0(\lam_k)  &   -\varpi^{(1)}_k\Delta_1(\lam_k) & \cdots &  (-1)^m\varpi^{(1)}_k\Delta_m(\lam_k)\\
\vdots & \vdots & \ddots & \vdots  \\
\varpi^{(m)}_k\Delta_0(\lam_k)  &  -\varpi^{(m)}_k\Delta_1(\lam_k) & \cdots &  (-1)^m\varpi^{(m)}_k\Delta_m(\lam_k)
\end{pmatrix}, \label{Tadjoint}
\end{align}
with
\begin{align}
 \hspace{-4mm} \nonumber \Delta_0(\lam_k)=
\begin{vmatrix}
\delta_{11}(\lam_k) & \cdots &  \delta_{1m}(\lam_k) \\
\vdots & \ddots & \vdots  \\
\delta_{m1}(\lam_k) & \cdots &  \delta_{mm}(\lam_k)
\end{vmatrix}, \quad
 \Delta_i(\lam_k)=
\begin{vmatrix}
\beta_1(\lam_k) & \cdots & \beta_m(\lam_k)\\
\delta_{11}(\lam_k) & \cdots &  \delta_{1m}(\lam_k) \\
\vdots & \ddots & \vdots  \\
\delta_{i-1,1}(\lam_k) & \cdots &  \delta_{i-1,m}(\lam_k) \\
\delta_{i+1,1}(\lam_k) & \cdots &  \delta_{i+1,m}(\lam_k) \\
\vdots & \ddots & \vdots  \\
\delta_{m1}(\lam_k) & \cdots &  \delta_{mm}(\lam_k)
\end{vmatrix} \quad (1\leq i\leq m).
\end{align}

Then, we expand $[T_x(\lam_k)+T(\lam_k)U(\lam_k)]T^*(\lam_k)$, $[T_y(\lam_k)+T(\lam_k)V(\lam_k)]T^*(\lam_k)$ and $[T_t(\lam_k)+T(\lam_k)$ $W(\lam_k)]T^*(\lam_k)$, obtaining that
\begin{eqnarray}
& &\hspace{-10mm} \nonumber  u_{1l}(\lam_k)= (-1)^{l-1}\Delta_{l-1}(\lam_k)\bigg\{\alpha_{x}(\lam_k)+\lam_k \alpha(\lam_k)+ \sum_{j=1}^m \beta_j(\lam_k)q_j  \\
& & \hspace{7mm} + \, \sum_{j=1}^m  \varpi^{(j)}_k \left[\beta_{j,x}(\lam_k) +\alpha(\lam_k)p_j -\lam_k \beta_j(\lam_k)\right]   \bigg\} \quad (1\leq l\leq m+1), \label{Kthrootsa} \\
& & \hspace{-10mm} \nonumber  u_{i+1,l}(\lam_k)= (-1)^{l-1}\Delta_{l-1}(\lam_k)\bigg\{\gamma_{i,x}(\lam_k)+\lam_k \gamma_i(\lam_k)+ \sum_{j=1}^m \delta_{ij}(\lam_k)q_j  \\
& & \hspace{10mm} +\, \sum_{j=1}^m\varpi^{(j)}_k \left[\delta_{ij,x}(\lam_k) +\gamma_i(\lam_k)p_j -\lam_k \delta_{ij}(\lam_k)\right]   \bigg\} \quad (1\leq i\leq m; \,1\leq l\leq m+1), \label{Kthrootsb} \\
& & \hspace{-10mm} \nonumber v_{1l}(\lam_k) =  (-1)^{l-1}\Delta_{l-1}(\lam_k) \bigg\{\alpha_y(\lam_k)  - \frac{1}{\sigma} \bigg[ \alpha(\lam_k) \bigg(2\,\lam_k^2  -  \sum_{n=1}^m p_n q_n\bigg)   -  \sum_{n=1}^m \beta_n(\lam_k) q_{n,x} \\
& & \hspace{6mm}  \nonumber + \, 2\,\lam_k \sum_{n=1}^m   \beta_n(\lam_k)q_n  \bigg]\bigg\}  +  (-1)^{l-1}\Delta_{l-1}(\lam_k) \sum_{j=1}^m \varpi^{(j)}_k
\bigg\{\beta_{j,y}(\lam_k)     \\
& &  \hspace{6mm} - \,\frac{1}{\sigma} \bigg[ \alpha(\lam_k) \left(2\,\lam_k p_j +p_{j,x}  \right) -  2\,\lam_k^2 \beta_j(\lam_k) +  \sum_{n=1}^m   \beta_n(\lam_k)q_n p_j    \bigg]\bigg\} \quad (1\leq l\leq m+1), \label{Kthrootsc} \\
& & \hspace{-10mm} \nonumber v_{i+1,l}(\lam_k) =(-1)^{l-1}\Delta_{l-1}(\lam_k) \bigg\{\gamma_{i,y}(\lam_k)  - \frac{1}{\sigma} \bigg[ \gamma_i(\lam_k) \bigg(2\,\lam_k^2  -  \sum_{n=1}^m p_n q_n\bigg) -  \sum_{n=1}^m \delta_{in}(\lam_k) q_{n,x} \\
 & & \hspace{10mm}   \nonumber + \, 2\,\lam_k \sum_{n=1}^m   \delta_{in}(\lam_k)q_n    \bigg]\bigg\} + (-1)^{l-1}\Delta_{l-1}(\lam_k)  \sum_{j=1}^m \varpi^{(j)}_k
\bigg\{\delta_{ij,y}(\lam_k) - \,\frac{1}{\sigma} \bigg[ -  2\,\lam_k^2 \delta_{ij}(\lam_k)   \\
& & \hspace{10mm} + \, \gamma_i(\lam_k) \left(2\,\lam_k\,p_j  + p_{j,x}  \right)  + \sum_{n=1}^m   \delta_{in}(\lam_k)q_n p_j   \bigg]\bigg\}  \quad (1\leq i\leq m; \,1\leq l\leq m+1), \label{Kthrootsd} \\
& & \hspace{-10mm} \nonumber w_{1l}(\lam_k) =  (-1)^{l-1}\Delta_{l-1}(\lam_k) \bigg\{\alpha_t(\lam_k)
+ \bigg[4\,\lam^3_k \alpha(\lam_k) - \alpha(\lam_k) \sum_{n=1}^m(2\,\lam_k p_nq_n + p_{n,x}q_n - p_{n}q_{n,x}  )   \\
& & \hspace{6mm}  \nonumber
+\sum_{n=1}^m \beta_n(\lam_k) \bigg(4\,\lam_k^2 q_n - 2\,\lam_k q_{n,x} -2\,q_n \sum_{r=1}^mp_rq_r + q_{n,xx} \bigg) \bigg] \bigg\} + (-1)^{l-1}\Delta_{l-1}(\lam_k)  \\
& & \hspace{6mm}  \nonumber \times\, \sum_{j=1}^m \varpi^{(j)}_k
\bigg\{\beta_{j,t}(\lam_k) + \bigg[\alpha(\lam_k)\bigg(4\,\lam_k^2p_j + 2\,\lam_k p_{j,x} - 2 p_j \sum_{n=1}^m p_n q_n  + p_{j,xx} \bigg)     \\
& & \hspace{6mm} -\,4\,\lam_k^3 \beta_{j}(\lam_k) + (2\,\lam_k p_j + p_{j,x}) \sum_{n=1}^m \beta_n(\lam_k)q_n - p_j \sum_{n=1}^m \beta_n(\lam_k)q_{n,x}     \bigg]\bigg\}  \, \, (1\leq l\leq m+1), \label{Kthrootse} \\
& & \hspace{-10mm} \nonumber w_{i+1,l}(\lam_k) =   (-1)^{l-1}\Delta_{l-1}(\lam_k) \bigg\{\gamma_{i,t}(\lam_k)
+ \bigg[4\,\lam^3_k \gamma_i(\lam_k) - \gamma_i(\lam_k) \sum_{n=1}^m(2\,\lam_k p_nq_n + p_{n,x}q_n - p_{n}q_{n,x}  )   \\
& & \hspace{2mm}  \nonumber
+\sum_{n=1}^m \delta_{in}(\lam_k) \bigg(4\,\lam_k^2 q_n - 2\,\lam_k q_{n,x} -2\,q_n \sum_{r=1}^mp_rq_r + q_{n,xx} \bigg) \bigg] \bigg\} +  (-1)^{l-1}\Delta_{l-1}(\lam_k)  \\
& & \hspace{2mm}  \nonumber \times\, \sum_{j=1}^m \varpi^{(j)}_k
\bigg\{\delta_{ij,t}(\lam_k) + \bigg[\gamma_i(\lam_k)\bigg(4\,\lam_k^2p_j + 2\,\lam_k p_{j,x} -2 p_j \sum_{n=1}^m p_nq_n  + p_{j,xx}\bigg)  -4\,\lam_k^3 \delta_{ij}(\lam_k)   \\
& & \hspace{2mm}  +\, (2\,\lam_k p_j + p_{j,x}) \sum_{n=1}^m \delta_{in}(\lam_k)q_n - p_j \sum_{n=1}^m \delta_{in}(\lam_k)q_{n,x}     \bigg]\bigg\}  \quad (1\leq i\leq m; \, 1\leq l\leq m+1).  \label{Kthrootsf}
\end{eqnarray}

Using Eqs.~\eref{UndetCoeffa} and recalling that $\Phi_k=(f_k, g^{(1)}_{k}, \dots, g^{(m)}_{k})^\top$ satisfies System~\eref{LaxZu} with $\lam=\lam_k$, we obtain
the derivatives of $\alpha(\lam_k)$ and $\gamma_{i}(\lam_k)$ with respect to $x$, $y$ and $t$ as follows:
\begin{eqnarray}
& & \hspace{-12mm}
\alpha_{x}(\lam_k) = -\sum_{j=1}^m \bigg(q_j -2\,\lam_k \varpi^{(j)}_k - \varpi^{(j)}_k\sum_{n=1}^{m}p_n\varpi^{(n)}_k\bigg) \beta_{j}(\lam_k)  -\sum_{j=1}^m \varpi^{(j)}_k \beta_{j,x}(\lam_k),
\label{UndetCoeffb}  \\
& &  \hspace{-12mm}
\gamma_{i,x}(\lam_k) = -\sum_{j=1}^m \bigg(q_j -2\,\lam_k \varpi^{(j)}_k - \varpi^{(j)}_k\sum_{n=1}^{m}p_n\varpi^{(n)}_k\bigg) \delta_{ij}(\lam_k)  -\sum_{j=1}^m \varpi^{(j)}_k \delta_{ij,x}(\lam_k)
\quad (1\leq i\leq m), \label{UndetCoeffc} \\
& &  \hspace{-12mm} \nonumber
\alpha_{y}(\lam_k) = \frac{1}{\sigma}\sum_{j=1}^m \bigg(2\,\lam_k q_j- q_{j,x} -4\,\lam^2_k \varpi^{(j)}_k + q_j\sum_{n=1}^{m}p_n\varpi^{(n)}_k + \varpi^{(j)}_k\sum_{n=1}^{m}p_n q_n  \\
& & \hspace{4mm} -2\,\lam_k\varpi^{(j)}_k \sum_{n=1}^m p_n \varpi^{(n)}_k - \varpi^{(j)}_k \sum_{n=1}^m p_{n,x} \varpi^{(n)}_k       \bigg) \beta_{j}(\lam_k)  -\sum_{j=1}^m \varpi^{(j)}_k \beta_{j,y}(\lam_k),
\label{UndetCoeffd}  \\
& &  \hspace{-13mm} \nonumber
\gamma_{i,y}(\lam_k) = \frac{1}{\sigma}\sum_{j=1}^m \bigg(2\,\lam_k q_j- q_{j,x} -4\,\lam^2_k \varpi^{(j)}_k + q_j\sum_{n=1}^{m}p_n\varpi^{(n)}_k + \varpi^{(j)}_k\sum_{n=1}^{m}p_n q_n  \\
& & \hspace{5mm} -2\,\lam_k\varpi^{(j)}_k \sum_{n=1}^m p_n \varpi^{(n)}_k - \varpi^{(j)}_k \sum_{n=1}^m p_{n,x} \varpi^{(n)}_k       \bigg) \delta_{ij}(\lam_k)  -\sum_{j=1}^m  \varpi^{(j)}_k \delta_{ij,y}(\lam_k) \, (1\leq i\leq m), \label{UndetCoeffe}
\\
& & \hspace{-12mm} \nonumber
\alpha_{t}(\lam_k) = -\sum_{j=1}^m\bigg\{ 4\,\lam_k^2 q_j - 2\,\lam_k q_{j,x} -2\,q_j \sum_{n=1}^mp_nq_n + q_{j,xx}-4\,\lam_k^3 \varpi^{(j)}_k + 2\,\lam_k q_j \sum_{n=1}^mp_n\varpi^{(n)}_k   \\
& & \nonumber \hspace{3mm}      - q_{j,x} \sum_{n=1}^mp_{n}\varpi^{(n)}_k + q_j\sum_{n=1}^m p_{n,x}\varpi^{(n)}_k  - \varpi^{(j)}_k\bigg[4\,\lam^3_k - \sum_{n=1}^m(2\,\lam_k p_nq_n + p_{n,x}q_n - p_{n}q_{n,x}  )  \\
& & \hspace{3mm} + \sum_{n=1}^m \varpi^{(n)}_k \bigg(4\,\lam_k^2p_n+2\,\lam_kp_{n,x} -2 \sum_{r=1}^m p_rq_r p_n+p_{n,xx} \bigg)       \bigg]    \bigg\} \beta_{j}(\lam_k)  -\sum_{j=1}^m \varpi^{(j)}_k \beta_{j,t}(\lam_k),
\label{UndetCoefff}  \\
& & \hspace{-12mm} \nonumber
\gamma_{i,t}(\lam_k) =  -\sum_{j=1}^m  \bigg\{ 4\,\lam_k^2 q_j - 2\,\lam_k q_{j,x} -2\,q_j \sum_{n=1}^mp_nq_n + q_{j,xx}-4\,\lam_k^3 \varpi^{(j)}_k + 2\,\lam_k q_j \sum_{n=1}^mp_n\varpi^{(n)}_k   \\
& & \nonumber \hspace{5mm}      -\, q_{j,x} \sum_{n=1}^mp_{n}\varpi^{(n)}_k + q_j\sum_{n=1}^m p_{n,x}\varpi^{(n)}_k  - \varpi^{(j)}_k\bigg[4\,\lam^3_k - \sum_{n=1}^m(2\,\lam_k p_nq_n + p_{n,x}q_n   \\
& &  \nonumber \hspace{5mm} - \, p_{n}q_{n,x}  ) + \sum_{n=1}^m \varpi^{(n)}_k \bigg(4\,\lam_k^2p_n+2\,\lam_kp_{n,x} -2 \sum_{r=1}^m p_rq_r p_n+p_{n,xx} \bigg)      \bigg]    \bigg\} \delta_{ij}(\lam_k) \\
& &  \hspace{5mm}  - \,\sum_{j=1}^m \varpi^{(j)}_k \delta_{ij,t}(\lam_k) \quad (1\leq i\leq m). \label{UndetCoeffg}
\end{eqnarray}

By means of Eqs.~\eref{UndetCoeffa} and \eref{UndetCoeffb}--\eref{UndetCoeffg}, we remove $\alpha(\lam_k)$, $\gamma_{i}(\lam_k)$, $\alpha_{x}(\lam_k)$, $\gamma_{i,x}(\lam_k)$, $\alpha_{y}(\lam_k)$, $\gamma_{i,y}(\lam_k)$, $\alpha_{t}(\lam_k)$ and $\gamma_{i,t}(\lam_k)$ from the right-hand sides of Eqs.~\eref{Kthrootsa}--\eref{Kthrootsf}, and finally prove that $u_{hl}(\lam_k)=0$, $v_{hl}(\lam_k)=0$ and $w_{hl}(\lam_k)=0$ ($1\leq h,l \leq m +1$).   \endproof

\end{appendix}

\newpage
\bibliographystyle{my-h-elsevier}

\end{document}